\documentclass[12pt]{iopart}

\usepackage{iopams}
\usepackage{setstack}
\usepackage{subfig}
\usepackage{graphicx}
\usepackage{float}
\usepackage{cite}
\usepackage[labelfont=bf, textfont=it]{caption}
\usepackage{hyperref}
\usepackage{xcolor}
\usepackage{tabularx}
\usepackage{colortbl}
\usepackage{multirow}
\usepackage{bm}
\usepackage{booktabs}
\usepackage{amssymb}
\usepackage{enumitem}
\usepackage{hhline}
\usepackage[normalem]{ulem}
\newcommand{\dd}{\mathrm d}

\newcommand{\vpp}{v_{\perp}}

\newcommand{\gradd}{\nabla}

\newcommand{\dv}{\mathrm{d}\mathbf{v}^{3}}

\newcommand{\dbp}{\delta \! B_{\parallel}}
\newcommand{\dap}{\delta \! A_{\parallel}}

\newcommand{\vds}{\mathbf{v}_{ds}}

\newcommand{\bb}{\mathbf{b}}
\newcommand{\B}{B}
\newcommand{\Bbold}{\mathbf{B}}
\definecolor{Gray}{gray}{0.85}
\definecolor{LightCyan}{rgb}{0.88,1,1}

\newcolumntype{a}{>{\columncolor{Gray}}c}

\begin{document}

\title[Electromagnetic instabilities in STEP]{On the importance of parallel magnetic-field fluctuations for electromagnetic instabilities in STEP}

\author{D. Kennedy$^1$, C. M. Roach$^1$, M. Giacomin$^{2,3},$ P. G. Ivanov$^4$, T. Adkins$^{5}$, F. Sheffield$^6,$ T. G{\"o}rler$^6$, A. Bokshi$^3$, D. Dickinson$^3$, H. G. Dudding$^{1},$ and B. S. Patel$^{1}$}

\address{$^1${UKAEA (United Kingdom Atomic Energy Authority),
Culham Campus,
Abingdon,
Oxfordshire,
OX14 3DB, UK.}}
\address{$^2$York Plasma Institute, University of York, York, YO10 5DD, United Kingdom}
\address{$^3$Dipartimento di Fisica ``G. Galilei'', Università degli Studi di Padova, Padova, Italy}
\address{$^4$Rudolf Peierls Centre for Theoretical Physics, University of Oxford, OX1 3PU, UK}
\address{$^5$Department of Physics, University of Otago, Dunedin, 9016, New Zealand}
\address{$^6$Max Planck Institute for Plasma Physics, Boltzmannstr. 2, Garching, 85748, Germany}

\ead{daniel.kennedy@ukaea.uk}

\begin{abstract}
    This paper discusses the importance of parallel perturbations of the magnetic-field in gyrokinetic simulations of electromagnetic instabilities and turbulence at mid-radius in the burning plasma phase of the conceptual high-$\beta$, reactor-scale, tight-aspect-ratio tokamak STEP.  Previous studies have revealed the presence of unstable \textit{hybrid} kinetic ballooning modes (hKBMs) and subdominant microtearing modes (MTMs) at binormal scales approaching the ion Larmor radius. In this STEP plasma it was found that the hKBM requires the inclusion of parallel magnetic-field perturbations to be linearly unstable. Here, the extent to which the inclusion of fluctuations in the parallel magnetic-field can be relaxed is explored through gyrokinetic simulations. In particular, the frequently used MHD approximation (dropping $\dbp$ and setting the $\nabla B$ drift frequency equal to the curvature drift frequency) is discussed and simulations explore whether this approximation is useful for modelling STEP plasmas. It is shown that the MHD approximation can reproduce some of the linear properties of the full STEP gyrokinetic system, but is too stable at low $k_y$ and nonlinear simulations using the MHD approximation result in very different transport states. 
    It is demonstrated that the MHD approximation is challenged by the high $\beta^{\prime}$ values in STEP, and that the approximation improves considerably at lower $\beta^{\prime}$. 
    Furthermore, it is shown that the sensitivity of STEP to $\dbp$ fluctuations is primarily because the plasma sits close to marginality and it is shown that in slightly more strongly driven conditions the hKBM \textbf{is} unstable without $\dbp.$  Crucially, it is demonstrated that the state of large transport typically predicted by local electromagnetic gyrokinetic simulations of STEP plasmas is not solely due to $\dbp$ physics. 

    \vspace{1ex}
    
    \end{abstract}

\section{Introduction}
\label{sec:intro}
The performance of magnetic-confinement-fusion devices such as spherical tokamaks (STs) is often limited by the presence of turbulent fluctuations that typically dominate the transport losses of heat, particles, and momentum. Therefore, understanding and predicting turbulent transport in next-generation STs is crucial for the optimisation of their performance. The UK STEP programme aims to generate net electric power $P_{\mathrm{el}} > 100\,$MW from fusion \cite{meyer2022,STEP}, by developing a compact prototype power plant, STEP, based on the ST concept. The first phase of this ambitious programme is to provide a conceptual design of a STEP prototype plant and reference plasma equilibria; in this context, a set of preferred STEP plasma scenarios are being designed \cite{tholerus2024}, using simplified reduced models to model the core transport. 

Reduced core-plasma transport models are essential for the integrated modelling of tokamak plasma scenarios, since first principles approaches are generally too computationally expensive to evaluate over confinement or resistive discharge timescales. Conventional-aspect-ratio devices tend to be dominated by electrostatic microturbulence driven by instabilities, such as modes driven by the ion-temperature gradient (ITG), modes driven by the electron-temperature gradient (ETG), and trapped-electron-driven modes (TEMs), and several reduced-complexity models have been developed to model electrostatic microturbulence (see e.g. \cite{tglf,qualikiz}). Such reduced models have been quite successful at predicting turbulent transport in conventional-aspect-ratio present-day devices at low-to-modest $\beta$ [where $\beta=2\mu_0p/B^2$ is the ratio of the total plasma pressure $p$ to the magnetic-field energy density $B^2/(2\mu_0)$, with $B$ the magnetic-field strength and $\mu_0$ the permeability of free space], where the turbulence is predominantly electrostatic: see e.g. \cite{Citrin_2022}. However, in STs, the confinement and turbulent transport, reviewed in \cite{kaye2021}, are both quantitatively and qualitatively different to that in conventional-aspect-ratio devices. ST geometry has favourable stability properties that allow access to higher-$\beta$ regimes where the turbulence is more electromagnetic in character and the dominant unstable modes include kinetic ballooning modes (KBMs) and microtearing modes (MTMs). These electromagnetic modes, less well understood than their electrostatic counterparts, are not well captured in the most advanced reduced core transport models; transport from electromagnetic turbulence has thus not been reliably captured in scenario modelling for STEP. 
 
Fortunately, most microinstabilities share broad characteristics which are well described by local\footnote{All simulations in this paper are local simulations; i.e. simulations are performed in a domain of perpendicular size that is infinitesimal in comparison with the length scale over which the equilibrium varies. Plasma equilibrium gradient {length scales} are taken to be constant across the simulation domain.} linearised $\delta \! f$ gyrokinetics (GK) provided that $k_{\perp}\rho_{s} \sim 1$ and $\rho_{\star s} \equiv \rho_s /a \ll 1$ where $\rho_s$ is the gyroradius of species $s,$ $a$ is a typical equilibrium length scale, and $k_\perp$ is the perpendicular wavenumber of the instability. High-fidelity GK simulations must be exploited: (i) to assess electromagnetic microinstabilities and their associated turbulence in the conceptual power-plant designs developed for STEP; and (ii) to improve the physics basis of reduced models used in scenario design.

In this paper, we concern ourselves with electromagnetic instabilities in STEP, in particular the hybrid-kinetic ballooning mode (hKBM) \cite{kennedy2023a} that GK simulations indicate to dominate turbulent transport \cite{giacomin2023b} in the conceptual designs of STEP. The inclusion of compressional magnetic fluctuations $\delta \! B_\parallel$ was found to be essential in \cite{kennedy2023a} for the hKBM to be unstable in STEP (see also discussion in Section \ref{sec:linear_GK}). However, $\delta \! B_\parallel$ physics is often simply missing from many higher-fidelity (e.g. global, full-$f$) GK codes and modelling tools. Instead, in GK theory and simulations of microinstabilities, it is common practice to neglect $\delta \! B_\parallel$, and to compensate for this by exploiting an approximate cancellation (valid in certain limits) that relies on the form of $\dbp$ in gyrokinetics and an exact relationship between the equilibrium $\gradd B$ and curvature drifts. In this work, we will refer to this compensation as the MHD approximation\footnote{In Section~\ref{sec:MHD_approx_implementation}, we will describe two separate implementations of the `MHD approximation' that have been included in some gyrokinetic codes. We will refer to these two different approximations as `MHD-1' and `MHD-2'. In \textbf{all} gyrokinetic simulations using the MHD approximation, $\dbp$ fluctuations are set to zero and a magnetic drift term is dropped to compensate for this.} (see e.g. \cite{berk_dominguez_1977,10.1063/1.873694,joiner2010}){; so called because it relies on (i) an MHD equilibrium force balance constraint on the relationship between the $\gradd B$ and curvature drifts, and (ii) the gyrokinetic form of $\dbp$ which excludes high frequency fast compressional Alfv\'en waves by enforcing perpendicular force balance on fluctuations $\delta \mathbf{j_\perp} \times \mathbf{B} = \mathbf{\nabla_{\perp} \cdot \delta p_\perp}$
}. The primary focus of this paper is to explore sensitivity of hKBM stability to the inclusion of $\delta \! B_{\parallel}$ and the appropriateness of the MHD approximation in STEP plasmas.

The paper is organised as follows: Section~\ref{sec:parallel_perturbations_intro_discussion} briefly reviews the results of \cite{kennedy2023a} and \cite{giacomin2023b} and outlines the motivation for using the MHD approximation in STEP plasmas. Following a brief review of electromagnetic $\delta \! f$ gyrokinetics, Section~\ref{sec:emgk} introduces how the MHD approximation fits into the GK framework and its implementation in GK codes. Section~\ref{sec:linear_GK} introduces the STEP equilibria and associated plasma parameters (with more details available in \cite{kennedy2023a,giacomin2023b}) and presents linear GK simulations (using the GK code \texttt{GENE}~\cite{GENE}) to assess the validity of the MHD approximation in STEP. It is explored whether the performance of the MHD approximation can be improved in conditions where the approximation's underlying assumptions are better satisfied. In Section~\ref{sec:nonlinear_gk} we demonstrate and explain why the MHD approximation is typically not appropriate for modelling electromagnetic turbulence in the STEP flat-top plasma. We again discuss cases where the MHD approximation can more faithfully model high$-\beta$ turbulence. Section~\ref{sec:parallel} explains that $\dbp$ is only required for instability in STEP because the local equilibrium is close to marginal stability (owing to substantial $\beta^\prime$ stabilisation). We demonstrate that it is in fact possible to drive electromagnetic hKBM turbulence in the absence of $\dbp$ in conditions further from marginal stability with or without implementing the MHD approximation on the magnetic drifts. The final summary and conclusions are presented in Section~\ref{sec:conclusions}. Additional auxiliary material is provided in Appendices, including a brief derivation of the MHD approximation that is applied in GK in \ref{sec:appendixA}, a discussion of the why {one of two proposed implementations of the MHD approximation is more physical} in \ref{sec:appendixb}, a useful expression for the magnetic-drift velocity given in \ref{sec:appendixc}, and turbulent snapshots in \ref{app:contours} from different simulations cited at various points in the main text.

\section{The importance of parallel magnetic-field perturbations}
\label{sec:parallel_perturbations_intro_discussion}

Recent GK analyses for mid-radius of the preferred flat-top operating point, STEP-EC-HD \cite{tholerus2024}, revealed the presence of unstable \textit{hybrid} KBMs\footnote{The term `\textit{hybrid} KBM' was introduced in \cite{kennedy2023a} to describe the dominant instabilities seen in STEP plasmas; typically these modes share attributes associated with KBM, ITG, and TEM.} (hKBM) and subdominant MTMs at binormal scales approaching the ion Larmor radius \cite{kennedy2023a,giacomin2023b} (note that similar MTMs were reported for an earlier proposed variant of STEP \cite{patel2021}). Nonlinear local GK simulations for the same local equilibrium in \cite{giacomin2023b} find that the hKBM turbulence can drive heat fluxes that exceed the available heating power by orders of magnitude in the absence of equilibrium flow shear, and that this large transport state is characterised by highly radially extended turbulent eddies. Further simulations in \cite{giacomin2023b} indicate that a transport steady state can be reached if equilibrium flow shear and/or $\beta^\prime$ stabilisation are sufficient at the flat-top to regulate the turbulence. (A transport steady state for a STEP flat-top was recently reported \cite{giacomin2024} using a physics-based model of transport from hKBM turbulence.) At lower $\beta^\prime$ and/or in the absence of equilibrium flow shear, however, hKBMs drive very large turbulent transport in all channels that may challenge access to the burning flat top through lower $\beta^{\prime}$ states. 

A natural extension to nonlinear simulations in \cite{giacomin2023b} is to attempt higher-fidelity \textbf{global} gyrokinetic
simulations for conditions where local GK predicts large turbulent fluxes. $\dbp$, indicated by local GK to be essential for the hKBM to be unstable in STEP-EC-HD \cite{kennedy2023a}, is unfortunately neglected in most global GK codes.
To progress towards a global description of hKBM-driven turbulence, we must both (i) implement $\dbp$ in a global-capable code;\footnote{Exciting progress
is being made in this direction (see e.g. \cite{mishchenko_borchardt_hatzky_kleiber_konies_nuhrenberg_xanthopoulos_roberg-clark_plunk_2023,YANG2023108892,wilms2023}), and there are several parallel efforts working towards implementing $\dbp$ globally in both Eulerian codes such as \texttt{GENE} \cite{GENE} and Particle-In-Cell codes such as \texttt{ORB5} \cite{orb5}.} and (ii) strive to better understand the sensitive dependence of the hybrid mode on $\dbp.$ This paper focuses on the latter.

\subsection{The MHD approximation for the magnetic-drift velocity}

As remarked in Section~\ref{sec:intro}, in GK theory and simulations of microinstabilities it is common to neglect the (typically destabilising\cite{aleynikova2018,Graves_2019})\footnote{It was demonstrated in \cite{Graves_2019} that neglecting $\dbp$ typically is stabilising for MHD-like modes because this omits the “magnetic compression” term in $\delta \! W$ (the potential energy of the plasma). If $\dbp$
is ignored then $\delta \! W$ is minimised incorrectly which has a stabilising effect on pressure-driven
modes.} parallel magnetic perturbation $\dbp$ at low $\beta,$ and to compensate for this by artificially adding a correction term proportional to $\nabla p$ to the magnetic drift velocity (see e.g. \cite{berk_dominguez_1977,bourdelle2003,joiner2010} and references therein). This idea of modifying the magnetic drifts, hereinafter referred to as the MHD approximation, is recommended in many GK codes on the neglect of $\dbp$. The merits, or lack thereof, of the MHD approximation have been tested in different GK simulations by various authors (see e.g. \cite{aleynikova2018}, \cite{joiner2010}). Whilst proper treatment of $\delta B_\parallel$ physics is often crucial \cite{aleynikova2018}, the cancellation of $\dbp$ and the $\nabla p$ term in the $\nabla{B}$ drift
has been demonstrated to be a good approximation when calculating the growth rates of KBMs at low $\beta$ and long wavelengths \cite{joiner2010}.  Here we explore whether this approximation may be appropriate for use in STEP plasmas\footnote{Section 3.3 of \cite{giacomin2023b} noted that the modes with longest perpendicular wavelength are typically expected to drive the most transport in STEP plasmas.}. 

To reiterate, the sensitivity of hKBM stability to $\dbp$ poses difficulties for  exploiting existing  codes for STEP. If the MHD approximation were valid for STEP, it would be useful for two key reasons: 

\begin{itemize}
    \item \textit{Global gyrokinetic codes often neglect} $\dbp.$

\vspace{1ex}

 Testing high turbulent fluxes predicted, but potentially not well resolved, by local GK in some local equilibria is of high priority for STEP design. Progress requires global gyrokinetic simulations including profile variation that should better resolve the box-scale streamers observed in local simulations \cite{giacomin2023b}.However, most global-capable codes simply do not include $\dbp$ at present { because of  the associated algorithmic complexity. The GK equation is closed by self-consistently solving the gyrokinetic form of 
Maxwell’s equations to evaluate the perturbed electromagnetic fields. Retaining the perpendicular part of Amp{\`e}re's law introduces a coupling between the electrostatic potential and the perpendicular magnetic potential $\dbp.$ Including $\dbp$ consistently requires the solution of a potentially poorly-conditioned system of equations involving $\delta \! \phi$ and $\dbp$, as well as derivatives and gyro-averages of these quantities. In a local code, these derivatives and gyro-averages are trivial to handle via spectral methods. In a global code, such an approach is not possible and one has to instead introduce complex ``gyrodisk averages'' to obtain a system which is numerically well behaved. (see \cite{YANG2023108892, wilms2023}).}
The MHD approximation has been used in a global code to study electromagnetic turbulence in NSTX plasmas \cite{Sharma2022}.

\item \textit{Integrated modelling tools often neglect} $\dbp$
\vspace{1ex}

Quasilinear turbulence models such as TGLF~\cite{tglf} provide fast transport predictions that are essential for the integrated modelling codes used in scenario design. These models perform a fast computation of the linear-mode properties, and feed these into saturation rules developed to parameterise the turbulent fluxes. As such, these reduced models rely on both the fidelity of the saturation rules, and the accuracy of the linear physics calculation, which requires capturing $\dbp$ effects in STEP.

\end{itemize}

Sections~\ref{sec:linear_GK}--\ref{sec:parallel} of this paper are devoted to: (i) exploring the degree to which the MHD approximation is appropriate for use in  STEP plasmas; and (ii) seeking to understand conditions where the requirement to include $\dbp$ fluctuations can be relaxed. 

\section{Linear electromagnetic gyrokinetics and the MHD approximation}
\label{sec:emgk}

We are interested in plasmas that are well described by the GK framework (see e.g., \cite{Abel_2013,catto_2019}): i.e. we are concerned with fluctuations $\sim e^{\mathrm{i} (\mathbf{k} \cdot \mathbf{r} -  \omega t)}$, having characteristic frequency $\omega$ and wavenumbers $k_\parallel$ and $k_\perp$ parallel and perpendicular to the equilibrium magnetic field direction $\bb = \Bbold/\B,$ that satisfy the standard GK ordering ${\omega}/{\Omega_{s}} \sim {\nu_{ss^{\prime}}}/{\Omega_{s}} \sim {k_\parallel}/{k_\perp} \sim {q_{s}\delta\phi}/{T_{0s}} \sim {\delta B_\parallel}/{\B} \sim {\delta \mathbf{B}_{\perp}}/{\B} \sim \rho_{\star,s} \ll 1,$ where $\Omega_{s} = q_{s}\B/m_{s}$ is the cyclotron frequency of species $s$ with charge $q_s,$ equilibrium density and temperature $n_{0s}$ and $T_{0s},$ respectively, mass $m_s$ and thermal speed $v_{\mathrm{th}s}=\sqrt{2T_{0s}/m_{s}}$, $\nu_{ss^{\prime}}$ is the typical collision frequency. The perturbed electrostatic potential is represented by $\delta\phi.$ Electromagnetic perturbations enter GK through $\dbp$ and $\delta \! B_{\perp},$ the fluctuations of the magnetic-field parallel and perpendicular to the equilibrium magnetic-field
direction. Electromagnetic effects are most conveniently described in local GK by writing the fluctuating magnetic field $\delta \! \mathbf{B} = \nabla \times (\delta \! \mathbf{A}_{\perp} + \delta\!  A_{\parallel} \bb) \simeq \nabla \times \delta \! \mathbf{A}_{\perp} + \nabla \delta \! A_{\parallel} \times \bb,$ then relating its parallel component to the fluctuating vector potential $\mathbf{A}$ by $\bb\cdot\delta \!\mathbf{B} \equiv \dbp \simeq \bb \cdot \nabla \times \delta \! \mathbf{A}_{\perp}.$

It is convenient to write the GK guiding-centre distribution function in the form $f_{s} = F_{0s}\left(1-{q_{s}\delta\!\phi}/{T_{0s}} \right) + g_{s} = F_{0s} + \delta \! f_{s}, \,\, \delta f_{s} = -({q_{s}\delta\!\phi}/{T_{0s}}) F_{0s} + g_{s}.$ Here, the GK distribution function consists of a Maxwellian piece $F_{0s}$ and an order-$\rho_{\star s}$-small perturbation $\delta \! f_{s},$ with $g_{s}$ being the non-adiabatic part of $\delta \! f_{s}.$ Under the above ordering, the collisionless, linear, GK equation is given in Fourier space by 
\begin{eqnarray}
    (\omega - \omega_{ds} - k_{\parallel}v_{\parallel})g_{s} &= \frac{q_{s}F_{0s}}{T_{0s}}(\omega - \omega_{\star s}^{T}) \nonumber \\  &\times J_{0}(b_{s})\left[ \delta \! \phi - v_{\parallel}\delta A_{\parallel} + \frac{m_{s}v_{\perp}^{2}}{q_{s}B} \frac{J_{1}(b_{s})}{b_{s}J_{0}(b_{s})}\delta B_{\parallel}\right]
    \label{eqn:linear_gk}
\end{eqnarray}
where: $b_{s} = k_\perp\vpp/\Omega_s$; $J_0$ and $J_1$ are Bessel functions of the first kind that arise from gyroaveraging; the drive frequency  $\omega_{\star s}^{T} = \omega_{\star s} \left[1 + \eta_{s} \left({v^{2}}/{v^{2}_{\mathrm{th}s}} - 3/2\right)\right]$ involves $\eta_{s} = \dd \ln T_{0s} / \dd n_{0s}$ and the diamagnetic frequency $\omega_{\star s} = ({k_{\perp}T_{0s}}/{q_{s}\B})\mathrm{d} \ln n_{0s} /\dd r$ is defined in terms of the radial coordinate $r$. The magnetic drift frequency is given by
\begin{equation}
    \omega_{ds} = \frac{1}{\Omega_{s}} \left(\omega_{\kappa} v_{\parallel}^{2} + \omega_{\nabla B} \frac{v_{\perp}^{2}}{2} \right)
    \label{eqn:drift_frequency}
\end{equation}
where the $\nabla B$ drift frequency $\propto \omega_{\nabla B} = \mathbf{k} \cdot \bb \times \nabla \B /\B$, and the curvature drift frequency $\propto \omega_{\kappa} = \mathbf{k} \cdot \bb \times (\bb \cdot \nabla \bb).$\footnote{Note that $\omega_{\kappa}$ and $\omega_{\nabla B}$ [appearing in (\ref{eqn:drift_frequency})] are \textbf{not} themselves frequencies. 
} The equilibrium magnetic field can be written as $\mathbf{B}_0 = \mathbf{\nabla} \psi \times \mathbf{\nabla} \alpha$, where $\psi$ is poloidal flux, and $\alpha=\xi - q(\psi)\theta - \nu(\psi, \theta)$ is the binormal coordinate,  with $\xi$ the toroidal angle, $\theta$ the poloidal angle, and $\nu(\psi, \theta)$  a periodic function of $\theta$ that depends on flux-surface shaping.  Perpendicular gradients can be expressed in terms of binormal and radial wavenumbers as $\mathbf{\nabla_{\perp}} = k_{\alpha}\mathbf{\nabla_{\alpha}} + k_\psi \mathbf{\nabla_{\psi}}$.

The fluctuating field quantities appearing in the GK equation are determined through the field equations. The perturbed electrostatic potential $\delta\!\phi$ is determined through quasineutrality
\begin{equation}
\sum_{s} \frac{n_{0s}q_{s}^{2}}{T_{0s}}\delta\!\phi = \sum_{s}q_{s}\int \dd^{3}\mathbf{v} \, J_{0}(b_{s}) g_{s}. 
\label{eqn:quasineutrality}
\end{equation}
The parallel magnetic vector potential $\delta \! A_{\parallel}$ is determined by parallel Amp{\`e}re's law
\begin{equation}
    k_{\perp}^{2}\delta \! A_{\parallel} = \mu_{0}\sum_{s} q_{s}\int \dd^{3} \mathbf{v} \, v_{\parallel} J_{0}(b_{s}) g_{s}.
    \label{eqn:ampere_parallel}
\end{equation}
The magnetic fluctuation $\dbp$ is determined by perpendicular Amp{\`e}re's law, leading to:
\begin{equation}
  \frac{\dbp}{\B} = -\frac{\mu_0}{\B^2} \sum_{s} \int \dv \,  g_{s} m_s v_{\perp}^2 \, \frac{J_{1}(b_{s})}{b_{s}}
     \label{eqn:ampere_perp}
\end{equation}
The MHD approximation is obtained in the long wavelength limit, where (\ref{eqn:ampere_perp}) reduces to:
\begin{equation}
    \lim_{b_s \to 0} \frac{\dbp}{\B} = -\frac{\mu_0}{2 \B^2} \sum_{s} \int \dv \, g_{s} m_s v_{\perp}^2 = -\frac{\mu_{0}}{\B^2} \delta P_{\perp} \label{eqn:pressure_balance}   
\end{equation}
involving the perturbed gyrokinetic perpendicular pressure that is defined in this limit as:
\begin{equation}
    \delta \! P_{\perp} = \sum_{s} \int \dv \, \frac{g_{s} m_s  v_{\perp}^2}{2}. \label{eqn:dbp}   
\end{equation}
Equation \ref{eqn:pressure_balance} implies that perpendicular force balance is satisfied to leading order in gyrokinetics, which is essential to exclude fast compressional Alfv\'en modes with frequencies $\sim \Omega/\sqrt{\beta}$ that lie outside of the gyrokinetic approximation (as discussed in Appendix\,A of \cite{roach2005}).

\subsection{The MHD approximation in GK}
\label{sec:the_MHD_approximation_in_GK}

The historical precedent for the MHD approximation in GK comes from the fact that, in the one fluid MHD
limit ($k_y\rho_s \rightarrow 0$), the compressional magnetic fluctuations cancel the contribution of the $\gradd B$ drift in a low-$\beta$ expansion in the absence of magnetic curvature \cite{berk_dominguez_1977}. Our first goal in this paper is to understand how this cancellation manifests itself in the GK framework. 

Linear GK force balance in ~\ref{eqn:pressure_balance} requires finite $\dbp$ to sustain a perpendicular pressure perturbation. It is therefore only appropriate to drop $\dbp$ if $\delta \! P_{\perp}$ can also be neglected.  In \ref{sec:appendixA} we calculate $\delta \! P_{\perp}$ from the perturbed gyrokinetic distribution function, exploiting an exact relationship (given in \ref{sec:appendixc}) between curvature and $\gradd B$ drifts that follows straightforwardly from equilibrium force balance:
\begin{equation}
    \omega_{\nabla B} = \omega_{\kappa} - (\mathbf{k} \cdot \bb) \frac{\gradd{p}}{\mu_0 \B^2},
\end{equation}
whereby $\omega_{\kappa}$ and $\omega_{\nabla B}$ differ by a term proportional to the pressure gradient.
It is demonstrated in \ref{sec:appendixA} that $\delta \! P_\perp$ vanishes in the long-wavelength, low-$\beta$ limit with $k_\parallel v_{\mathrm{th}s} \ll \omega_{ds} \ll \omega$, if: 
\begin{equation}
    \omega_\kappa - \omega_{\nabla B} = 0.
    \label{eq:MHD_approx}
\end{equation}
Enforcing equation~\ref{eq:MHD_approx} can therefore be used to implement the MHD approximation and drop $\dbp$ from GK in the appropriate limits. 

Two approaches have been adopted to enforce equation~\ref{eq:MHD_approx} in gyrokinetics.  Denoting the physical values (i.e. those set by the actual equilibrium magnetic geometry) of $\omega_{\kappa}$ and $\omega_{\gradd B}$  by $\omega_{\kappa}^{\mathrm{act}}$ and $\omega_{\gradd B}^{\mathrm{act}},$ we can satisfy \ref{eq:MHD_approx} by either:

\begin{itemize}
    \item (MHD-1) {setting the $\gradd B$ drift equal to the curvature drift} \[ (\omega_{\kappa},\omega_{\gradd B}) = (\omega_{\kappa}^{\mathrm{act}},\omega_{\kappa}^{\mathrm{act}}) \] 
    \item (MHD-2) {setting the curvature drift equal to the $\gradd B$ drift} \[ (\omega_{\kappa},\omega_{\gradd B}) = (\omega_{\gradd B}^{\mathrm{act}},\omega_{\gradd B}^{\mathrm{act}}). \]
\end{itemize}
Both have been implemented in GK codes. \ref{sec:appendixb} explains why MHD-1 is better motivated physically for the study of a particular equilibrium, which is consistent with arguments given in Appendix~A of \cite{Zocco2015MagneticCA} and Appendix~F.5 of \cite{garbet:hal-03974985}.

\subsection{Implementation of the MHD approximation in gyrokinetic codes}
\label{sec:MHD_approx_implementation}
  
It is helpful to connect the MHD approximation to the magnetic-drift velocity $\mathbf{v}_{d}$ rather than the magnetic-drift frequency $\omega_{ds} = \mathbf{k}_{\perp} \cdot \mathbf{v}_{ds}$. The exact magnetic drift velocity is given by 
\begin{eqnarray}
     \vds &= \bb \times \left( \frac{v_{\parallel}^{2}}{\Omega_s} (\bb\cdot\gradd)\bb + \frac{\vpp^2 }{2\Omega_s} \nabla \ln \B \right)\label{eqn:full_drift}
 \end{eqnarray}
 where it is shown in \ref{sec:appendixc} that:
 \begin{eqnarray}
    \gradd_{\perp} \ln \B &= (\bb\cdot \gradd)\bb - \frac{\mu_0 \gradd p}{\B^2}.  
\end{eqnarray}
We can now understand both implementations of the MHD approximation in terms of adding or subtracting corrections to this definition. 

\subsubsection{The MHD-1 approximation} The historical precedent for setting the $\nabla B$ drift equal to the curvature drift comes from a result of \cite{10.1063/1.873694} showing that, in the one fluid MHD limit ($k_y\rho_D \rightarrow 0$), there is a cancellation of terms such that $\dbp$ can be dropped as long as the $\nabla B$ drift is corrected by a term proportional to $\nabla p$ so that it is equal in magnitude to the curvature drift and points in the same direction. This will be referred to as ``MHD-1'': 
\begin{eqnarray}
     \vds &= \bb \times \frac{\bb\cdot\gradd \bb}{\Omega_s} \left( v_{\parallel}^{2}+\frac{\vpp^2 }{2} \right) \label{eqn:gradb_eq_curv}
 \end{eqnarray}
In MHD-1 a term proportional to the pressure gradient has been added to the right hand side of Equation~\ref{eqn:full_drift}.

\subsubsection{The MHD-2 approximation}
In MHD-2 the drift velocity is written as:
\begin{eqnarray}
     \vds &= \bb \times \frac{\gradd \B}{\B\Omega_s} \left( v_{\parallel}^{2}+\frac{\vpp^2 }{2} \right) \label{eqn:curv_eq_gradb}
 \end{eqnarray}
 As discussed in \ref{sec:appendixc}, this effectively corresponds to changing the equilibrium magnetic field curvature by setting $\bb\cdot\gradd \bb = \gradd \B/\B$, and again removing the pressure gradient contribution to the $\nabla B$ drift in this different local equilibrium.  MHD-2 allows $B_{\parallel}$ to be dropped from the GK equation (\ref{eqn:linear_gk}), but modifies the local equilibrium.

{MHD-1 is the more physical implementation because it simply exploits a cancellation between two terms {\em without changing the local equilibrium curvature}; henceforth where we use the term ``MHD approximation'' and this will imply MHD-1.}\footnote{Both MHD-1 and MHD-2 are implemented in the GENE code with MHD-1 as the default setting.}

\section{Linear gyrokinetic simulations using the MHD approximation}
\label{sec:linear_GK}

The gyrokinetic analysis in this paper focuses primarily on a single equilibrium flux-surface taken from close to mid-radius ($q = 3.5, \Psi_n = 0.49$) in the STEP reference scenario STEP-EC-HD \footnote[1]{SimDB UUID: 2bb77572-d832-11ec-b2e3-679f5f37cafe, Alias: smars/jetto/step/88888/apr2922/seq-1} (where EC stands for Electron Cyclotron heating and current drive, and HD stands for High Density), which is designed to deliver a fusion power $P_{\mathrm{fus}} = 1.8$~GW. We note that this is precisely the same flux surface as examined in \cite{giacomin2023b}, a choice that has been made intentionally to allow benchmarking between our results against those reported in \cite{giacomin2023b}. It should also be noted that whereas \cite{giacomin2023b} focused on accurate investigation of the turbulent transport in STEP; the focus of this paper is instead on a more conceptual study of the hKBM. 

A Miller parameterisation \cite{miller1998} was
used to model the local magnetic equilibrium geometry, and the shaping parameters
were fitted to the chosen surface using \texttt{pyrokinetics} ~\cite{pyrokinetics2024}, a Python library developed to facilitate pre- and post-processing of gyrokinetic analysis performed using a range of different GK codes. \texttt{pyrokinetics} also contains an ideal-ballooning solver which has been used throughout this work to ensure that all equilibria remain MHD-stable when the local equilibrium parameters are varied. Table \ref{tab:table_1} provides the values of various local equilibrium quantities at the flux surface examined in this paper, including magnetic shear $\hat{s}$; safety factor $q$; normalised minor radius $\rho/a$; elongation $\kappa$ and its radial derivative $\kappa^\prime$; triangularity $\delta$ and its radial derivative $\delta^\prime$; the radial derivative of the Shafranov shift $\Delta^\prime$; and the normalised inverse density and temperature gradient scale lengths of species $s$, $a/L_{ns}$ and $a/L_{Ts},$ respectively. Our simulations evolve three species: electrons, deuterium, and tritium and neglect entirely impurities and fast particles. The interested reader is referred to \cite{kennedy2023a,giacomin2023b} for more details on the equilibrium and on the setup of the computational grids, which are identical to those used in the aforementioned works.

\begin{table}[htbp]
\centering
\begin{tabular}{l c l c }
\hhline{~~~}
\hhline{-|-|-|-}
Parameter & Value & Parameter & Value \\
$\Psi_n$ & 0.49 & $\rho/a$ & 0.64  \\
$q$ & 3.5 & $\hat{s}$ & 1.2\\
$\beta_e$ & 0.09 &  $\beta'$ & -0.48 \\
$\kappa$ & 2.56 & $\kappa'$ & 0.06 \\
$\delta$ & 0.29 & $\delta^\prime$ & 0.46 \\
$k_y^{n=1}\rho_D$ & 0.0047 & $\Delta'$ & -0.40 \\
$a/L_{n_e}$ & 1.03 &  $a/L_{T_e}$ & 1.58 \\
 $a/L_{n_D}$ &  1.06 & $a/L_{T_D}$ & 1.82 \\
$a/L_{n_T}$ & 0.99 & $a/L_{T_T}$ &  1.82 \\
\hhline{-|-|-|-}
\end{tabular}
\caption{{Local equilibrium quantities at the STEP mid-radius flux surface considered in this paper.}}
\label{tab:table_1}
\end{table}

\begin{figure}[htbp]
    \centering
    \subfloat[]{\includegraphics[width=0.48\textwidth]{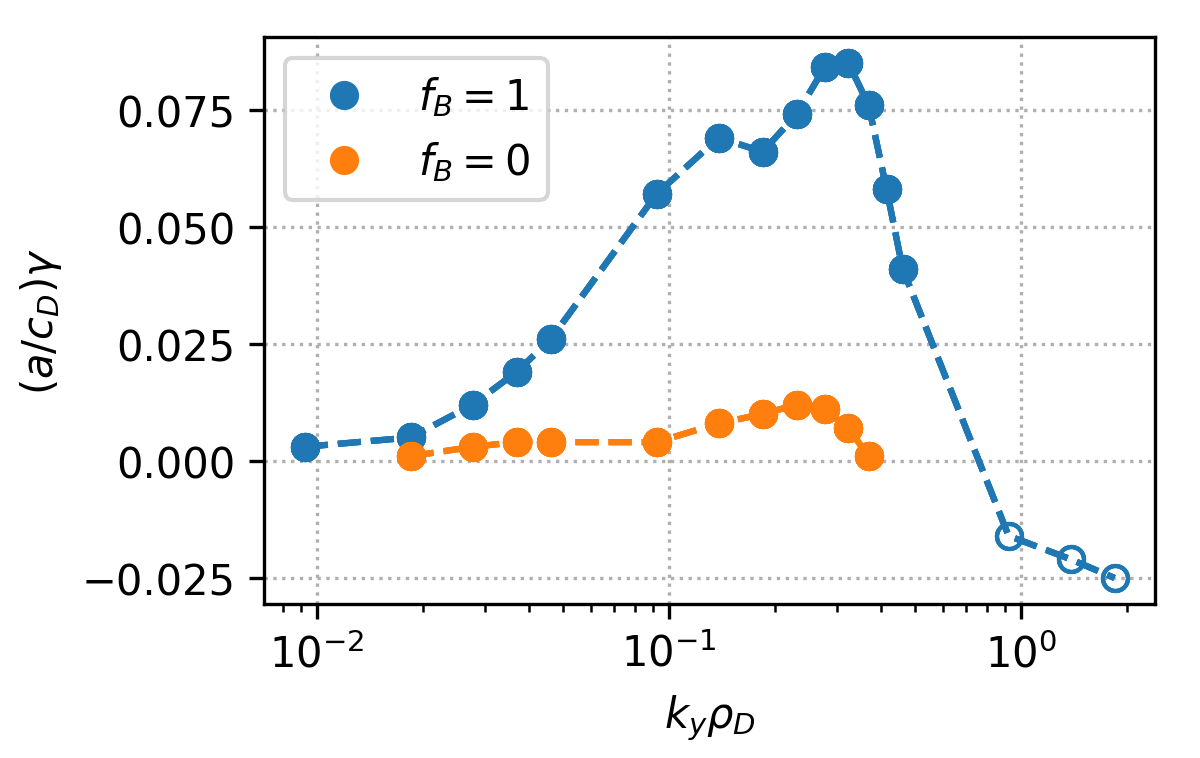}}\quad
    \subfloat[]{\includegraphics[width=0.48\textwidth]{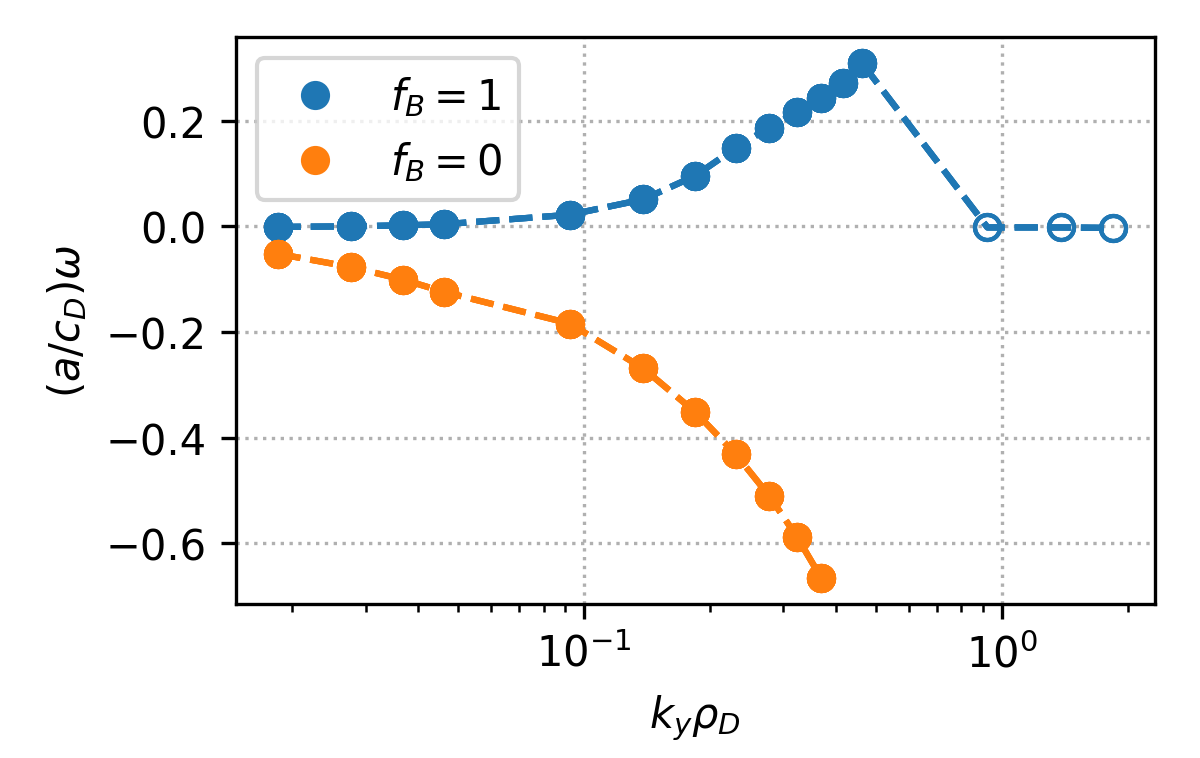}}
    \captionsetup{font=it}
    \caption{Growth rate \emph{(\textbf{a})} and mode frequency \emph{(\textbf{b})} as functions of the binormal wavenumber from linear simulations of the dominant instability in STEP-EC-HD on a mid-radius flux surface. Simulations are shown both with, $f_B=1$ (blue), and without, $f_B=0$ (orange), $\delta B_\parallel$. The two simulations are otherwise identical and the exact magnetic-drift velocity is used in each simulation.}
    \label{fig:bpar_comp}
\end{figure}

Previous linear analysis shows that the hKBM is the dominant ion-scale instability on this surface, with a subdominant MTM also found to be unstable on a subset of these binormal scales (see Fig 19 and Fig 20 of \cite{kennedy2023a}). No unstable microinstabilities are observed at the electron Larmor radius scale. The dominant hKBM and the subdominant MTM can both be recovered physically; that is, one can recover the subdominant mode by either by exploiting the up-down symmetry in the local equilibrium and forcing the parity of the perturbed distribution function in an initial-value calculation, or by using an eigenvalue solver to return the unstable linear spectrum. However, importantly for our work, it was also shown that for this equilibrium it is possible to recover the subdominant mode simply by artificially suppressing $\delta B_\parallel$ (thus stabilising the hKBM). This latter point is exemplified in Fig~\ref{fig:bpar_comp}.

Fig~\ref{fig:bpar_comp} shows the linear growth rate (a) and frequency (b) (normalised to the deuterium sound speed, $c_D= \sqrt{T_e/m_D}$, divided by the minor radius of the last closed flux surface) as functions of the binormal wavenumber $k_y\rho_D=n\rho_{\ast D}d\rho/d\Psi_n$. Table~\ref{tab:table_1} also includes the binormal wavenumber $k_y^{n=1}\rho_D$ corresponding to the toroidal mode number $n = 1$. Simulations are shown both including, $f_B=1$ (blue), and neglecting, $f_B=0$ (orange), $\delta \! B_\parallel$ fluctuations (the two simulations are otherwise identical) and no approximations are made concerning the magnetic drift velocity. Importantly, we see from Fig~\ref{fig:bpar_comp} that if $\delta \! B_\parallel$ is artificially excluded from calculations (as routinely assumed in a number codes and modelling tools) then we recover the previously subdominant MTM (note the change in frequency) as the fastest growing unstable mode in the system. Succinctly, the hKBM is linearly stable on this surface (along with many other surfaces in the STEP equilibrium) without $\delta \! B_\parallel$.

\subsection{Linear gyrokinetic simulations using the MHD approximation in STEP}
\label{subsec:MHD_appropriate_linear}

The MHD approximation introduced in Section~\ref{sec:the_MHD_approximation_in_GK} may allow us to study the hKBM in the absence of $\dbp$ fluctuations. In Section~\ref{sec:the_MHD_approximation_in_GK}, we introduced two implementations: (i) MHD-1 ($\mathbf{v}_{d}$ given by equation~\ref{eqn:gradb_eq_curv}); and (ii) MHD-2 ($\mathbf{v}_{d}$ given by equation~\ref{eqn:curv_eq_gradb}), that were designed to compensate for the effect of excluding $\dbp.$ We now seek to explore whether either of these approximations are appropriate for modelling STEP plasmas. 

\begin{figure}[htbp]
    \centering
    \subfloat[]{\includegraphics[width=0.48\textwidth]{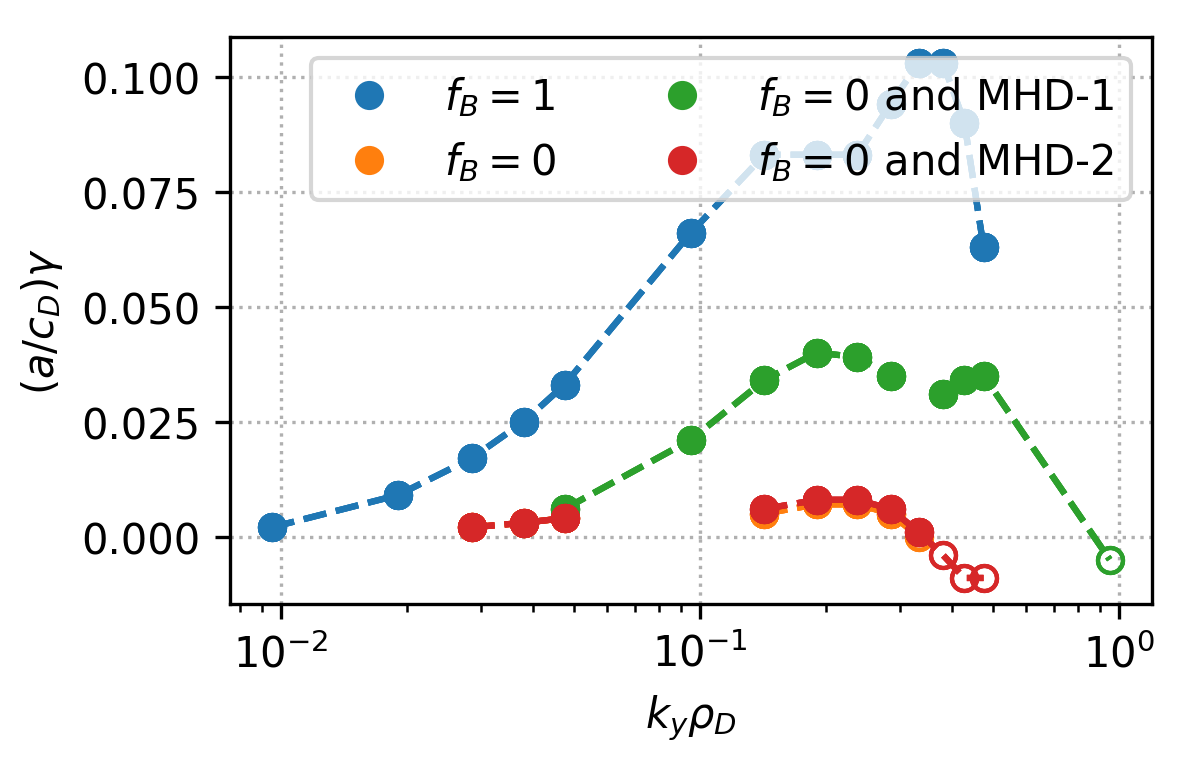}}\quad
    \subfloat[]{\includegraphics[width=0.48\textwidth]{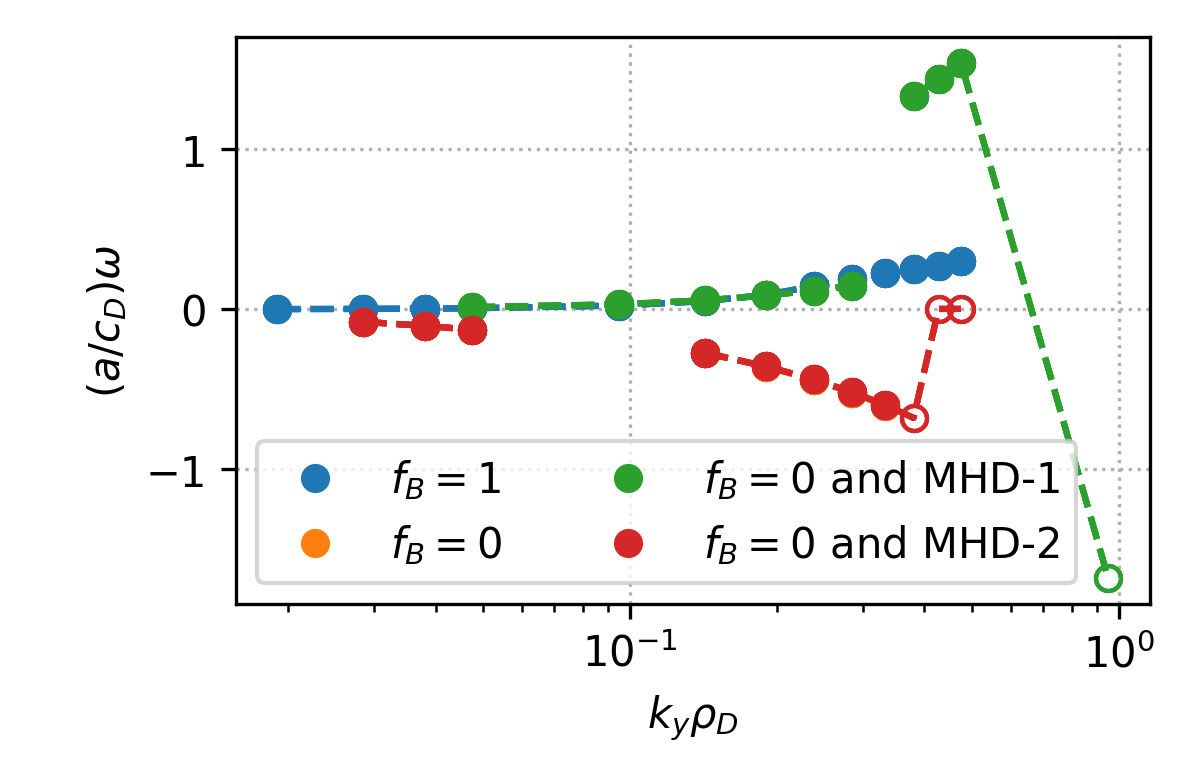}}
    \captionsetup{font=it}
     \caption{Growth rate \emph{(\textbf{a})} and mode frequency \emph{(\textbf{b})} as functions of the binormal wavenumber from linear simulations of the dominant instability in STEP-EC-HD on a mid-radius flux surface. Simulations are shown both with, $f_B=1$ (blue), and without, $f_B=0$ (orange, green, red), $\delta  
     \! B_\parallel$. For the simulations 
     without $\dbp,$ different treatments of the drift velocity are also shown: (i) the full drift velocity [$\mathbf{v}_{ds}$ given by equation~\ref{eqn:full_drift} (orange)]; (ii) MHD-1 [$\mathbf{v}_{ds}$ given by equation~\ref{eqn:gradb_eq_curv} (green)]; and MHD-2 [$\mathbf{v}_{ds}$ given by equation~\ref{eqn:curv_eq_gradb} (red)]. Note that the MHD-2 curve (red) lies on top of the curve that excludes $\dbp$ but uses the exact form of the magnetic-drift velocity (orange). }
    \label{fig:bpar_comp2}
\end{figure}
Fig~\ref{fig:bpar_comp2} shows the linear growth rate (a) and frequency (b) as functions of the binormal wavenumber for simulations both including, $f_B=1$ (blue), and excluding, $f_B=0$ (orange, green, red) $\dbp$ fluctuations. The simulations without $\dbp$ fluctuations employ different prescriptions of the magnetic drift velocity: (i) the full drift velocity [$\mathbf{v}_{d}$ given by equation~\ref{eqn:full_drift} (orange)]; (ii) MHD-1 [$\mathbf{v}_{d}$ given by equation~\ref{eqn:gradb_eq_curv} (green)]; and MHD-2 [$\mathbf{v}_{d}$ given by equation~\ref{eqn:curv_eq_gradb} (red)]. Using the MHD-2 approximation (Fig~\ref{fig:bpar_comp2}, red) we see that the hKBM is linearly stable and we once again recover the previously subdominant MTM (note the change in frequency) as the fastest growing unstable mode in the system.\footnote{In fact, the simulation using the MHD-2 approximation (red) returns the exact same linear spectrum as the simulation that excludes $\dbp$ but uses the exact form of the magnetic-drift velocity (orange). This arises because (i) the hKBM is more stable at reduced curvature in the MHD-2 local equilibrium, and (ii) MTMs are typically only very weakly sensitive to $\gradd p$ (see Fig~15b of \cite{patel2021}).} However, using the MHD-1 approximation [setting the $\nabla B$ drift parallel to the curvature drift (Fig~\ref{fig:bpar_comp2}, green)] does find the hKBM but with a strongly reduced growth rate. These results indicate that the MHD-1 approximation is the more appropriate treatment of the magnetic drift when $\dbp$ is neglected in STEP plasmas, as we might expect from the fact that MHD-1 does not change the local equilibrium. However, we note that even though this approximation recovers the hKBM, accurately capturing this mode clearly requires a proper treatment of $\dbp$ and exact magnetic drifts (particularly at low $k_y\rho_D$ where MHD-1 suggests hKBMs are stable at wavenumbers where full physics finds the mode to be unstable). 

\subsection{Improving the fidelity of the MHD approximation}
\label{subsec:improving_linear_fidelity}

This section explores whether it is possible to study the hKBM in the absence of $\dbp$ by modifying the local equilibrium parameters such that the MHD approximation is better satisfied. There are three key assumptions in the derivation of the MHD approximation presented in \ref{sec:appendixA}: 

\begin{itemize}
    \item[(i)] A long-wavelength assumption $k_y\rho_D \ll 1.$
    \item[(ii)] An MHD-like ordering of the frequencies $k_{\parallel}v_{\mathrm{th}s} \ll \omega_{ds} \ll \omega. $
    \item[(iii)] A low-$\beta$ approximation $\beta \ll 1.$
\end{itemize}These assumptions are explored in the following subsections through parameter scans. 

\subsubsection{Assumption (i) \texorpdfstring{$k_y \rho_D \ll 1$}{TEXT}} 

The long-wavelength assumption in STEP is well satisfied since we are primarily concerned with modes with $k_{y}\rho_D \ll 1.$ 
 
\subsubsection{Assumption (ii) \texorpdfstring{$k_\parallel v_{\mathrm{th}s} \ll \omega_{ds} \ll \omega$ and Assumption (iii) $\beta \ll 1$}{TEXT}} \label{sec:betaprime_discussion} 

Here we show that the difficulty in improving the fidelity of the MHD approximation lies in satisfying Assumption (ii) and Assumption (iii) simultaneously.  

\begin{figure}
    \centering
  \subfloat[]{\includegraphics[width=0.48\textwidth]{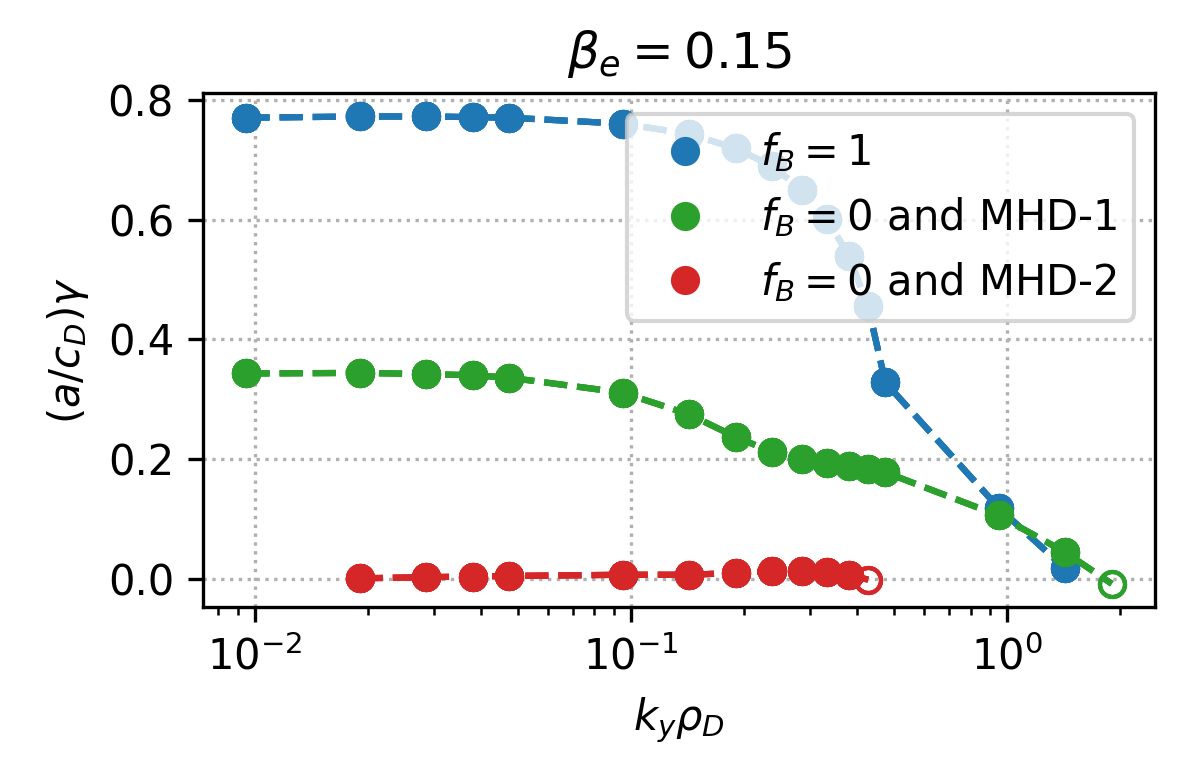}}\quad
   \subfloat[]{\includegraphics[width=0.48\textwidth]{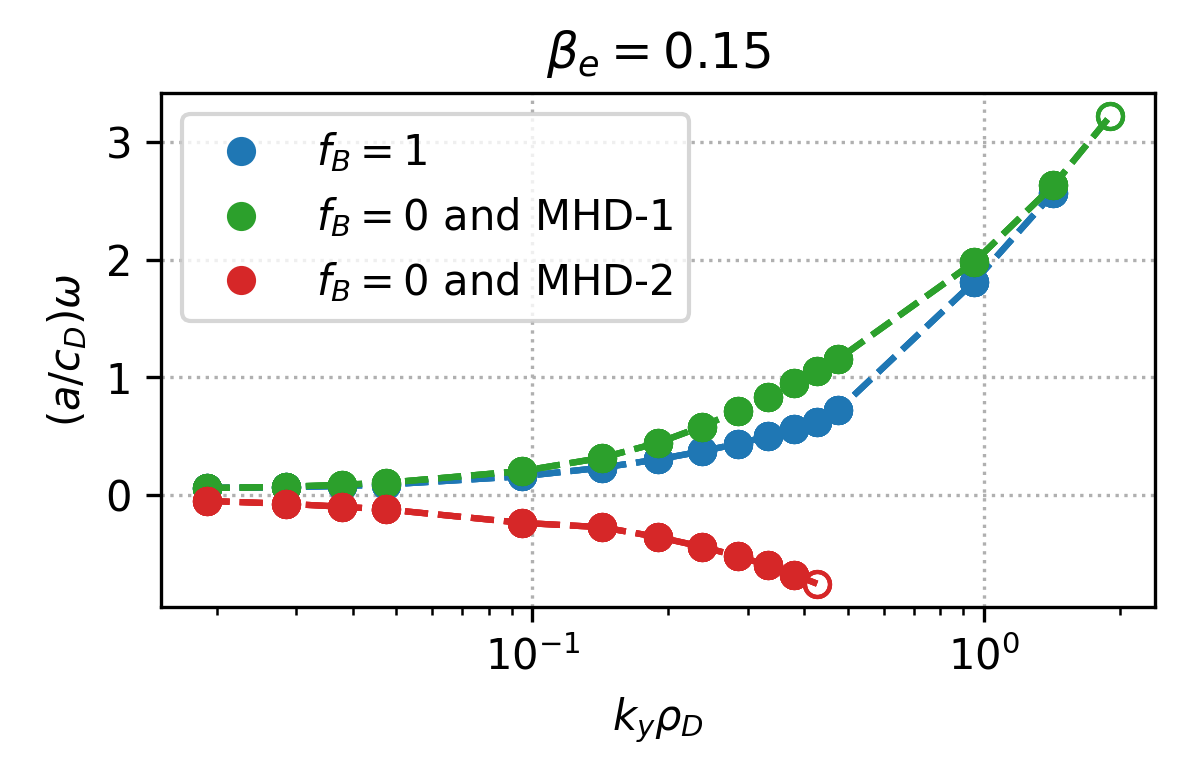}}\\
  \subfloat[]{\includegraphics[width=0.48\textwidth]{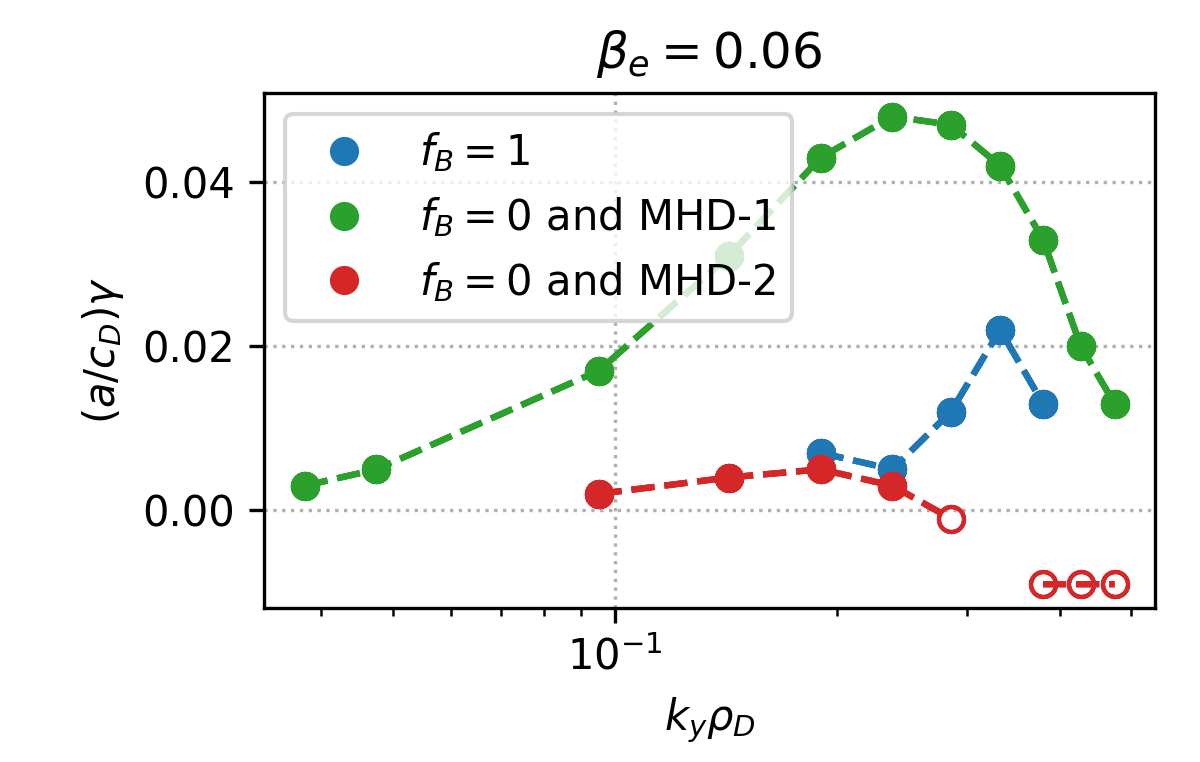}}\quad
   \subfloat[]{\includegraphics[width=0.48\textwidth]{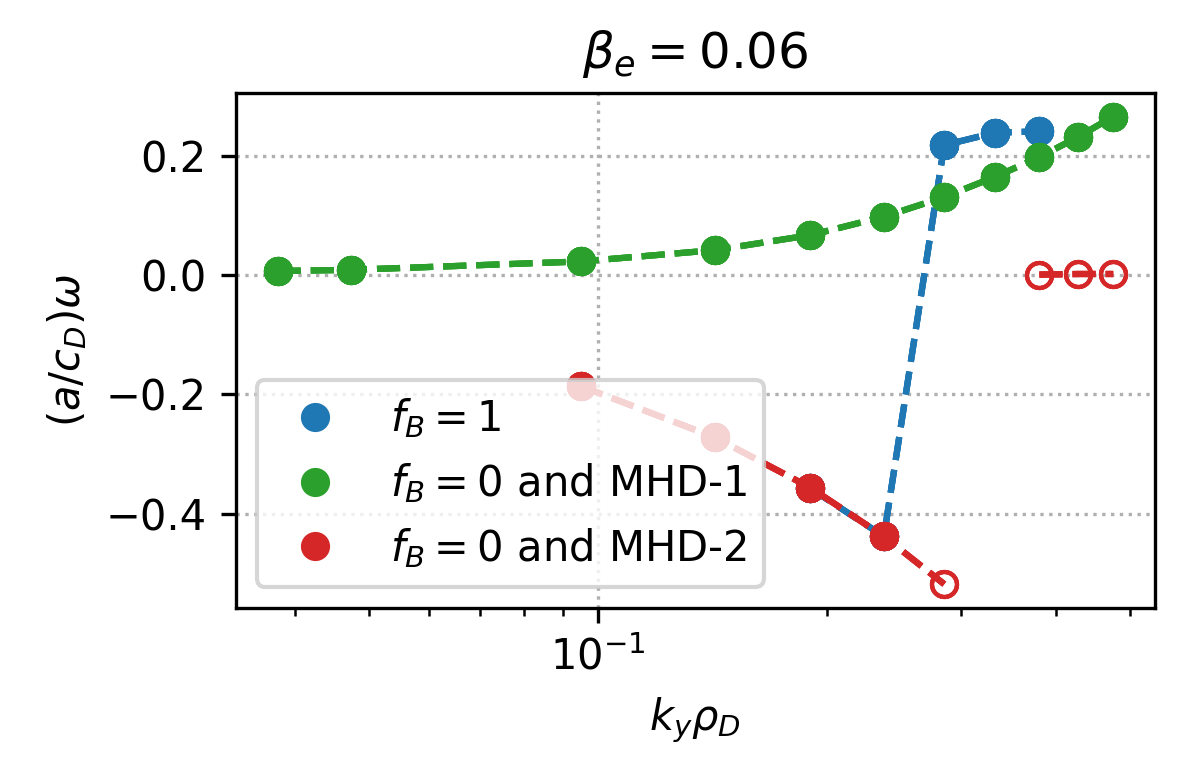}} \\ 
  \subfloat[]{\includegraphics[width=0.48\textwidth]{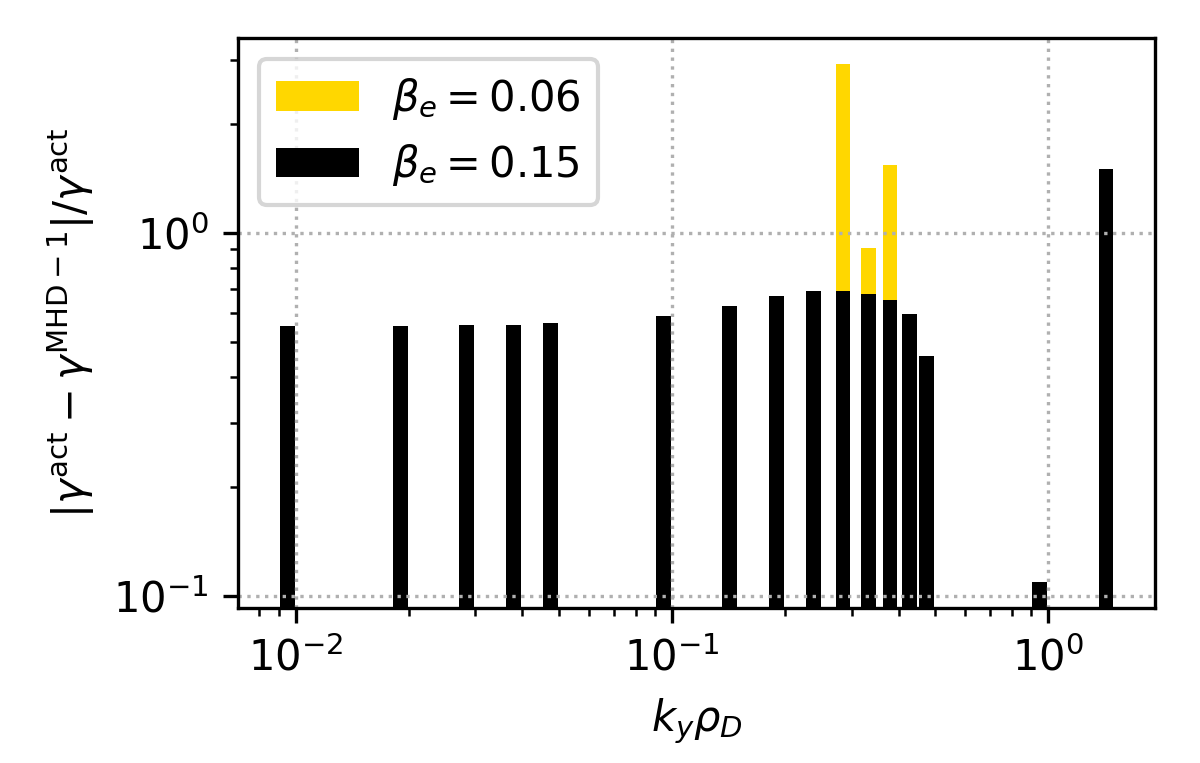}}\quad
   \subfloat[]{\includegraphics[width=0.48\textwidth]{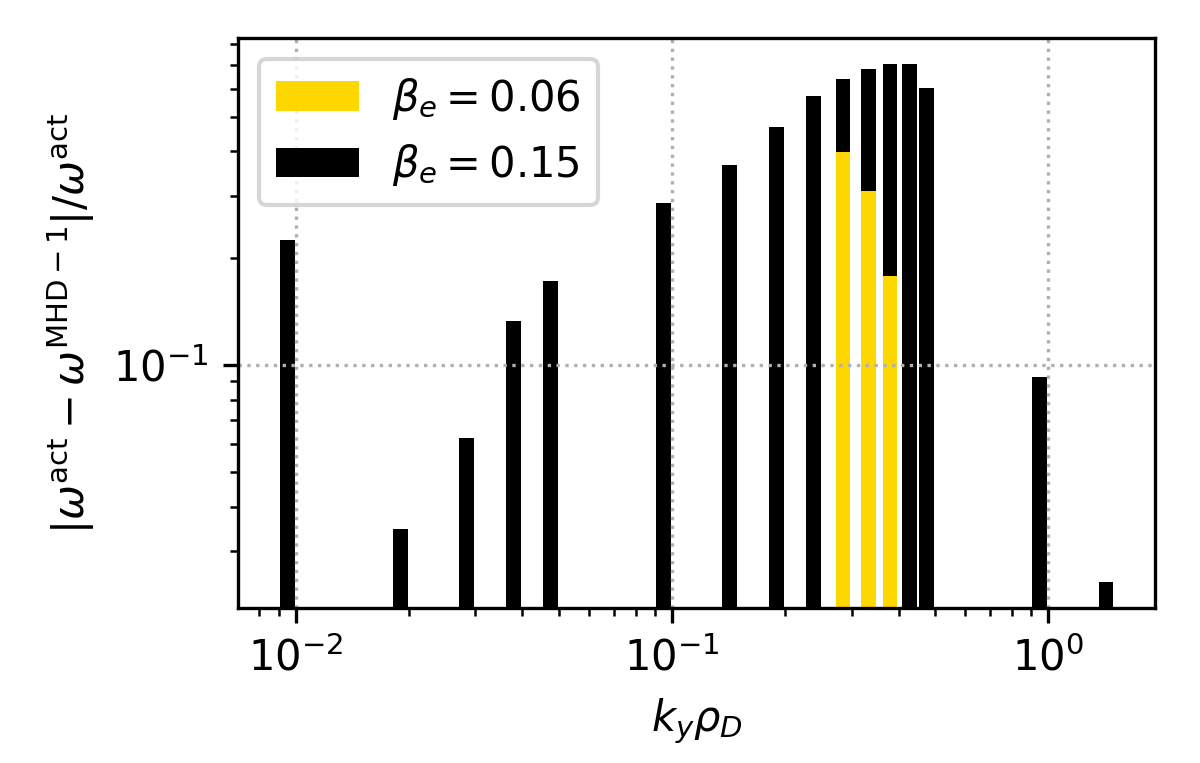}}
    \caption{Growth rates \emph{(\textbf{a, c})} and mode frequencies \emph{(\textbf{b, d})} as functions of the binormal wavenumber from linear simulations of the dominant instability in STEP-EC-HD on a mid-radius flux surface. The value of $\beta_e$ has been varied keeping $\beta^\prime$ fixed (all other local equilibrium quantities are the same as in Table~\ref{tab:table_1}). Results are shown for $\beta_e = 0.15$ \emph{(\textbf{a, b})} and $\beta_e = 0.06$ \emph{(\textbf{c, d})}. Simulations are shown both with, $f_B=1$ (blue), and without, $f_B=0$ (green, red), $\delta \! B_\parallel$. For the simulations without $\dbp,$ different treatments of the drift velocity are also shown: (i) MHD-1 (green); and (ii) MHD-2 (red). The absolute error made by the MHD-1 approximation for the growth rate and the mode frequency are shown in panels \emph{(\textbf{e})} and \emph{(\textbf{f})} respectively. The error is computed only for modes which are unstable and propagating in the ion-direction.}
    \label{fig:b_scan}
\end{figure}

The argument leading to the MHD approximation, given in \ref{sec:appendixA} and \ref{sec:appendixb}, requires $\delta \! P_{\perp}/(n_s T_s) \ll q_s \delta \phi/(n_s T_s)$.  From Equation~\ref{eqn:B6}: 
\[ \delta \! P_{\perp}  \propto (\omega_{\mathrm{dia}}/\omega)  \ \delta \! \phi, \]  
it follows that we require:
\[ \omega_{\mathrm{dia}} \ll \omega. \]
Now for an Alfv{\'e}nic instability $\omega\sim k_{\parallel} v_{ti}/\sqrt{\beta}$, and with $\omega_{\mathrm{dia}}\sim k_{\perp} v_{\mathrm{dia}} \sim \rho_{\ast}k_{\perp} v_{ti} \beta^{\prime}/\beta$ and $k_{\parallel} \sim \rho_{\ast} k_{\perp}$, this in turn implies:  
\begin{equation}
    \beta^{\prime} \ll \sqrt{\beta} \label{eq:betaprime}.
\end{equation}   
Equation~\ref{eq:betaprime}, together with $\beta \ll 1$, describes the parameter regime where we expect the MHD approximation to hold, and shows that reducing $\beta$ on its own (i.e. at fixed $\beta^\prime$) can make satisfying (\ref{eq:betaprime}) more difficult! Fig~\ref{fig:b_scan} shows the results of simulations using different values of $\beta$ (at fixed $\beta^\prime$) and both including, $f_B=1$ (blue), and excluding, $f_B=0$ (green, red), $\dbp$ fluctuations. Simulations without $\dbp$ fluctuations employ either the MHD-1 (green) or MHD-2 (red) approximation. As expected, reducing $\beta$ at fixed $\beta^{\prime}$ gives no noticeable improvement in the performance of the MHD approximation. 

On the other hand, equation~\ref{eq:betaprime} shows that reducing $\beta^\prime$ at fixed $\beta$ should always improve the fidelity of the MHD approximation, as will be discussed further in Section~\ref{sec:parallel}.

\section{Nonlinear GK simulations using the MHD approximation (MHD-1)}
\label{sec:nonlinear_gk}

  {From here we proceed using the more physical implementation of the MHD approximation, MHD-1, which still captures hKBMs as the fastest growing linear instability in STEP, albeit with reduced growth rates especially at low $k_y$ (see Section~\ref{subsec:MHD_appropriate_linear}).}
This suggests that it might be possible to simulate hKBM driven turbulence in STEP using the MHD approximation in a global code, where the full inclusion of $\dbp$ is not yet routinely available. As a first step towards this goal, we perform a \textit{local} nonlinear simulation using the MHD approximation; this simulation is otherwise identical to the hKBM simulation performed in \cite{giacomin2023b} in the absence of equilibrium flow shear.

\subsection{The MHD approximation does not accurately reproduce the large-transport state predicted by full-physics simulations}
\label{sec:MHD_step?}
\begin{figure}[htbp]
    \centering

  \subfloat[]{\includegraphics[width=0.48\textwidth]{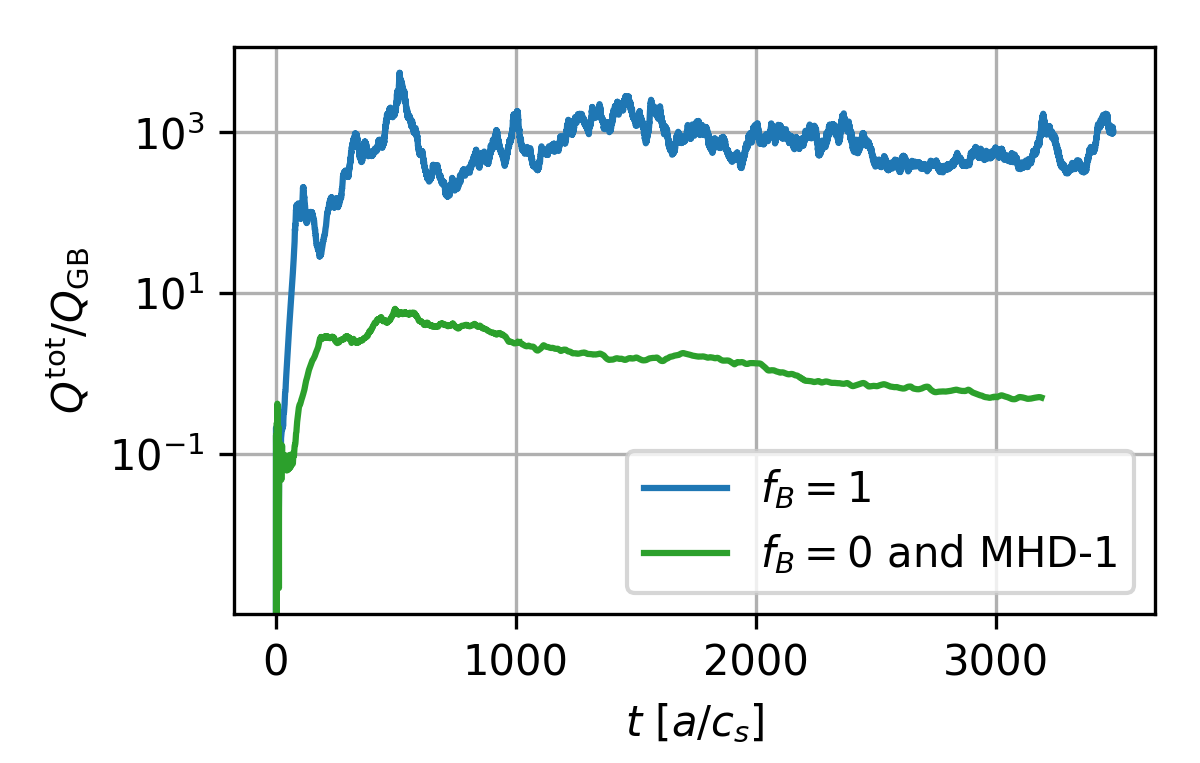}}\quad
   \subfloat[]{\includegraphics[width=0.48\textwidth]{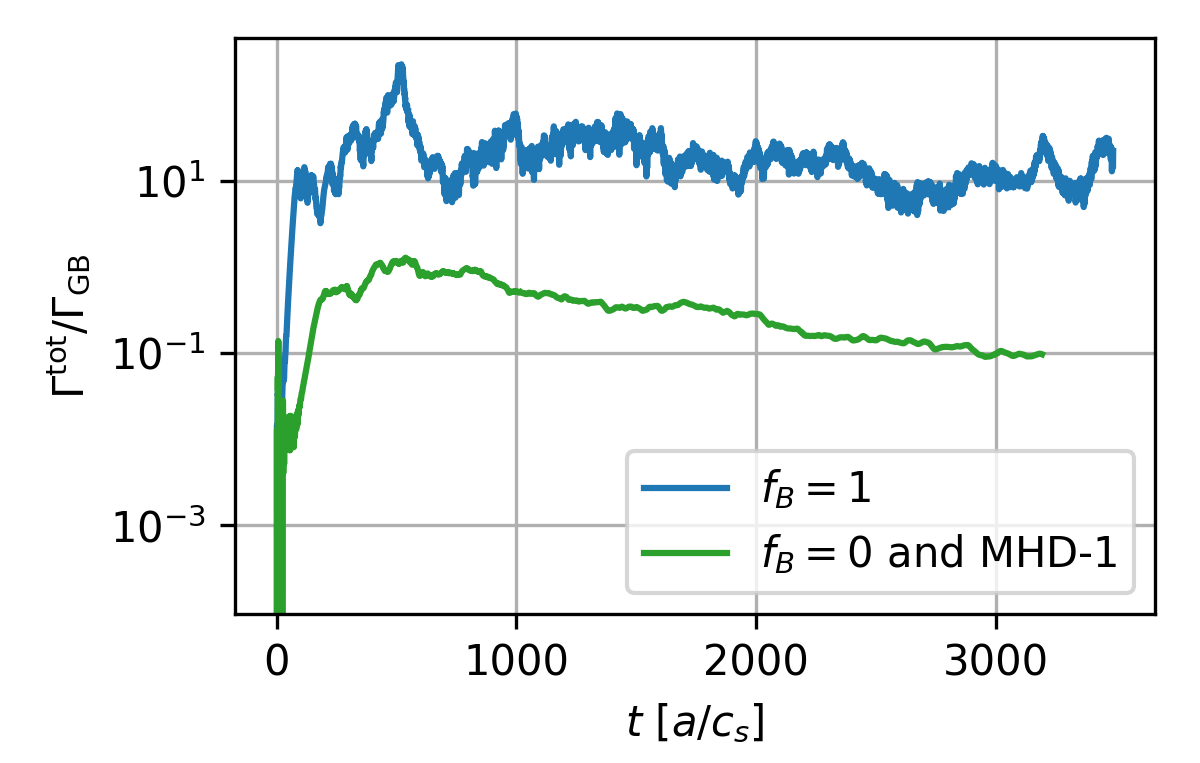}}\\
  \subfloat[]{\includegraphics[width=0.48\textwidth]{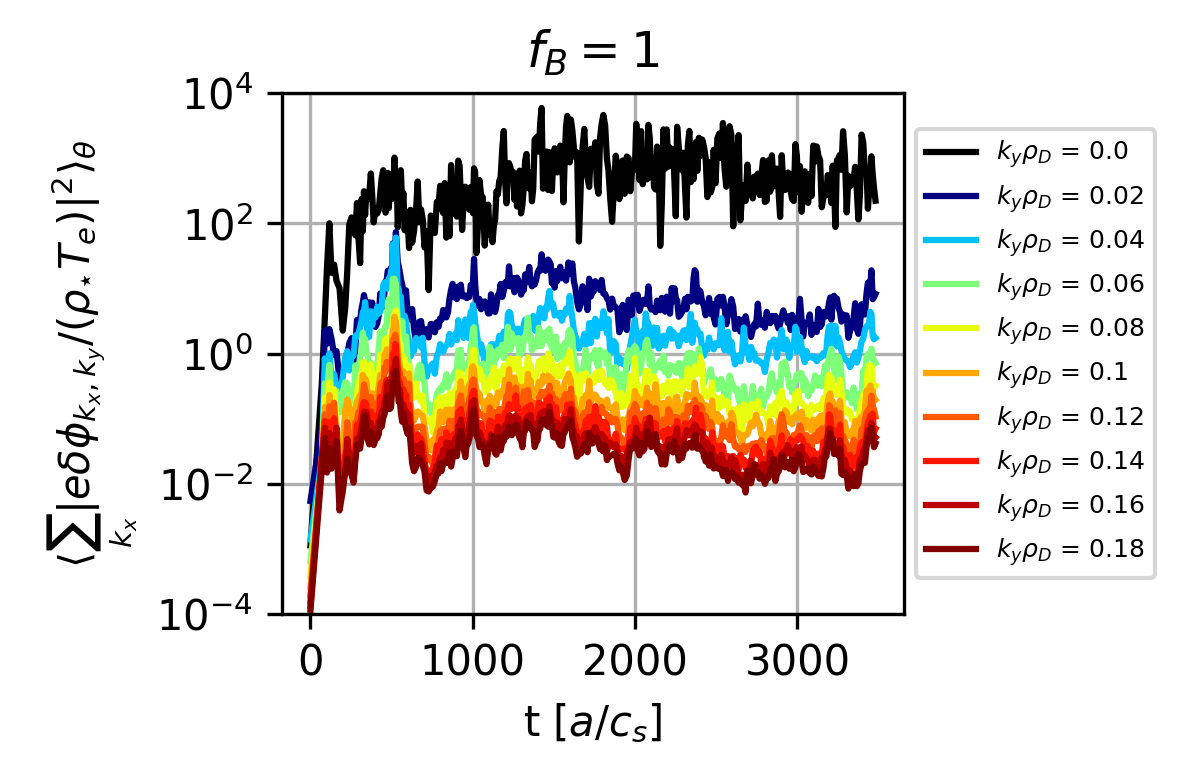}}\quad
   \subfloat[]{\includegraphics[width=0.48\textwidth]{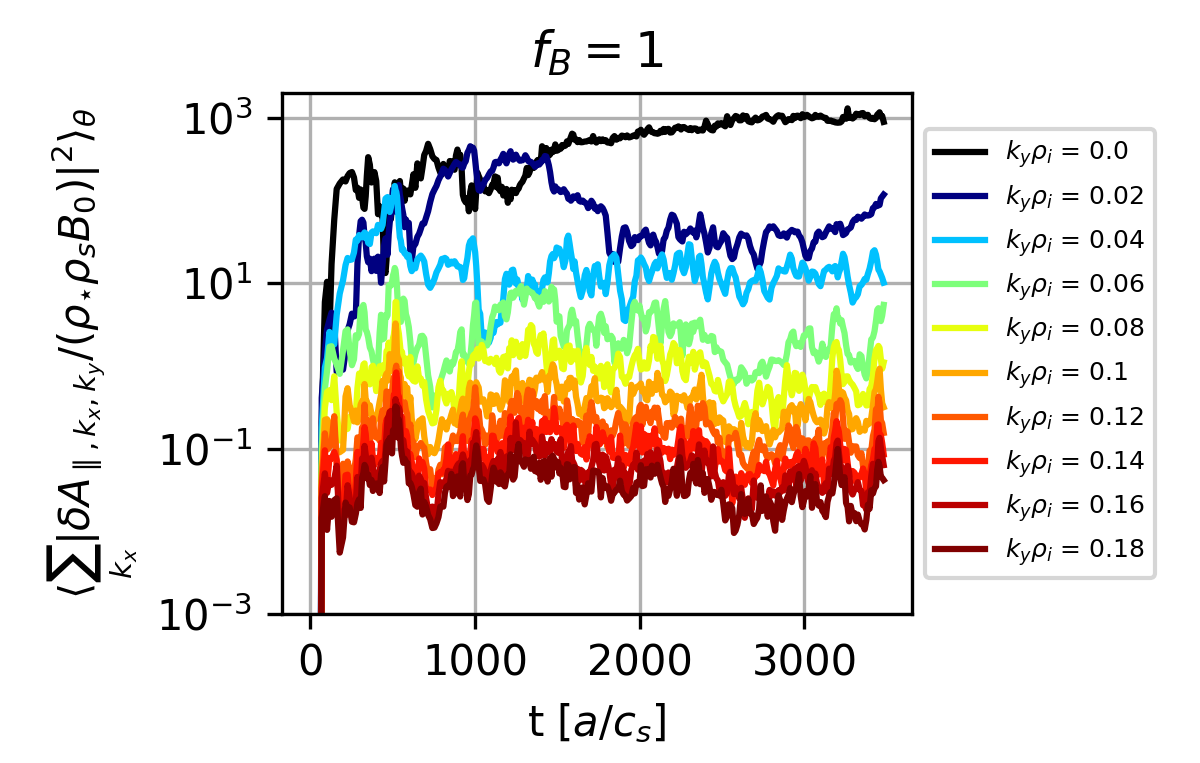}} \\ 
   \subfloat[]{\includegraphics[width=0.48\textwidth]{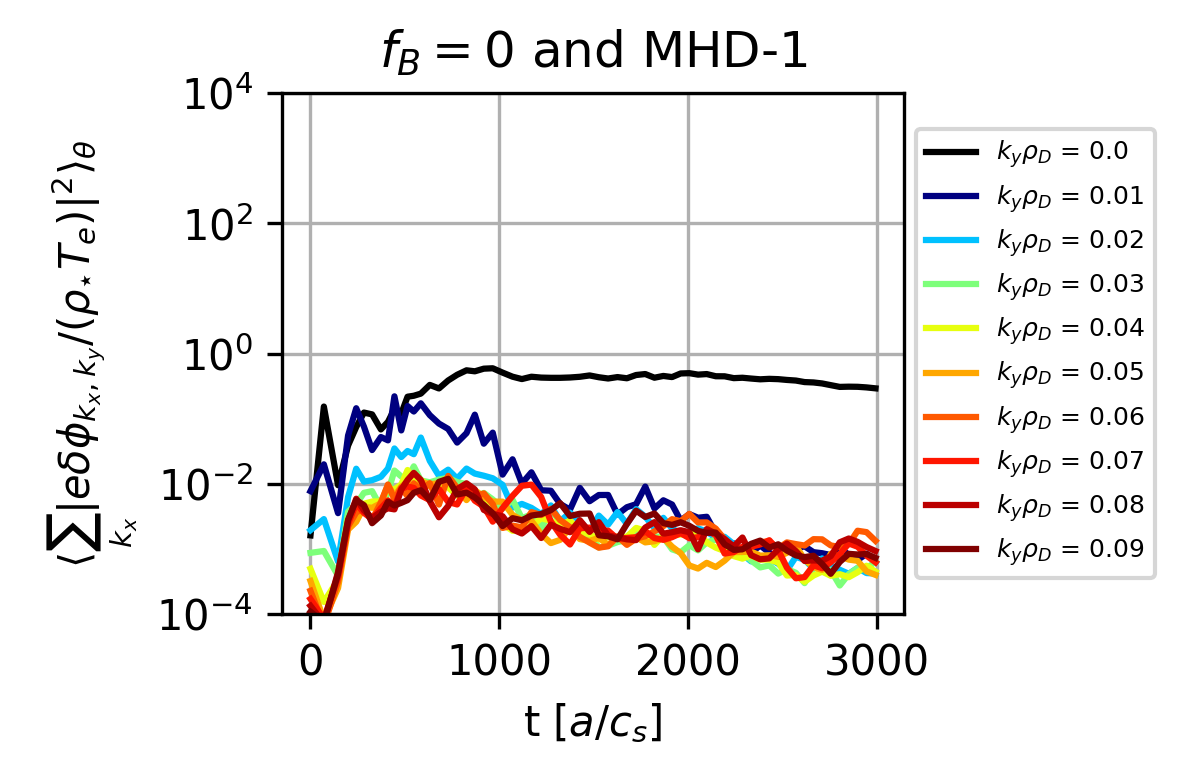}}\quad
    \subfloat[]{\includegraphics[width=0.48\textwidth]{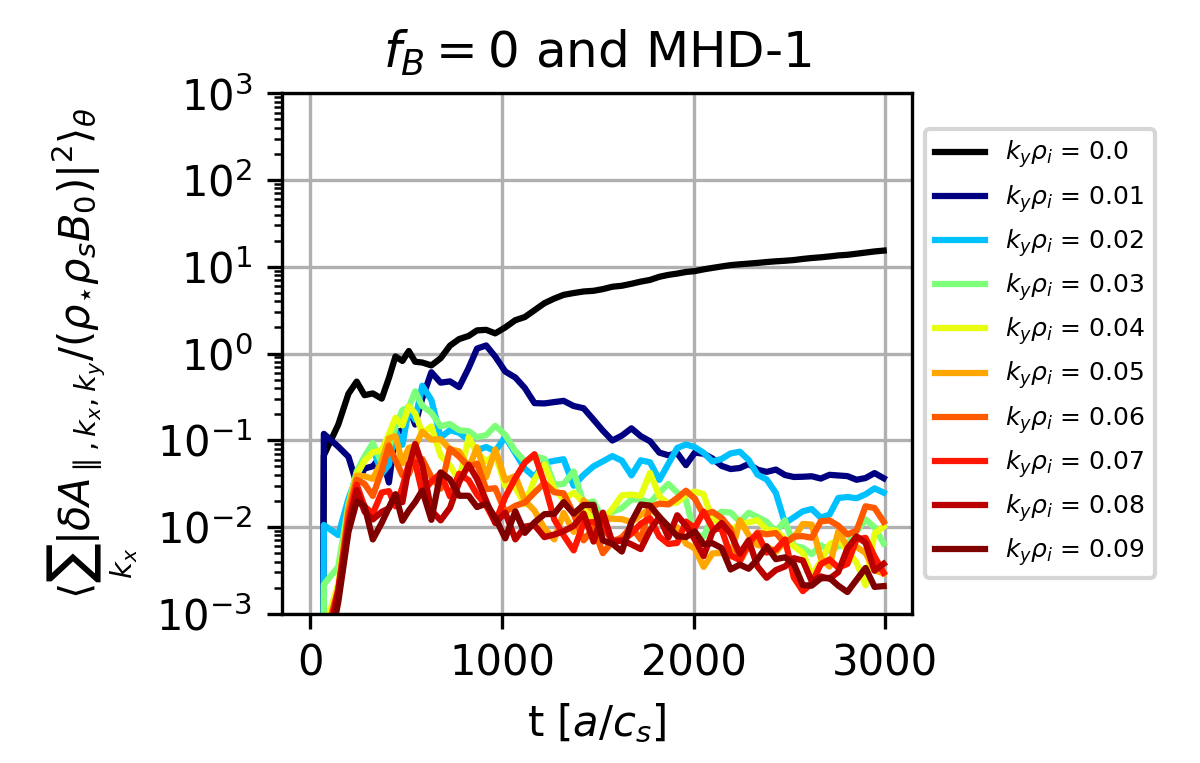}}
    
    \captionsetup{font=it}
    \caption{Time traces of the total heat flux \emph{(\textbf{a})} and total particle flux \emph{(\textbf{b})} from \texttt{GENE} simulations. Simulations are shown both with, $f_B=1$ (blue), and without, $f_B=0$ (green), $\delta B_\parallel$. The case without $\dbp$ fluctuations uses the MHD approximation (MHD-1) with $\mathbf{v}_{d}$ given by equation~\ref{eqn:gradb_eq_curv}. Also shown is the time trace of the $k_y$ spectrum of $\delta \! \phi$ \emph{(\textbf{c}, \textbf{e})} and $\delta \! A_\parallel$ \emph{(\textbf{d}, \textbf{f})}  from the same simulations. Only the first decade of poloidal wavenumbers is shown.}
    \label{fig:bpar_comp_sec5}
\end{figure}

\begin{figure}
\centering
   \subfloat[]{\includegraphics[width=0.48\textwidth]{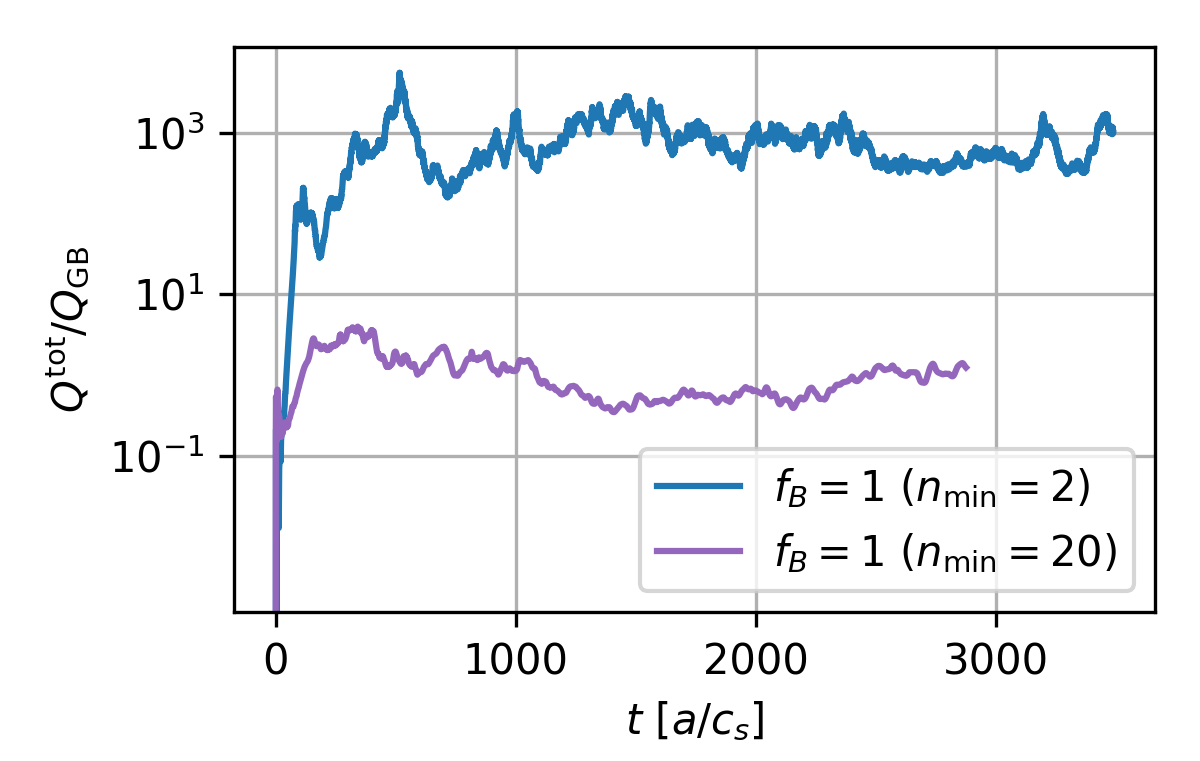}}\quad
    \subfloat[]{\includegraphics[width=0.48\textwidth]{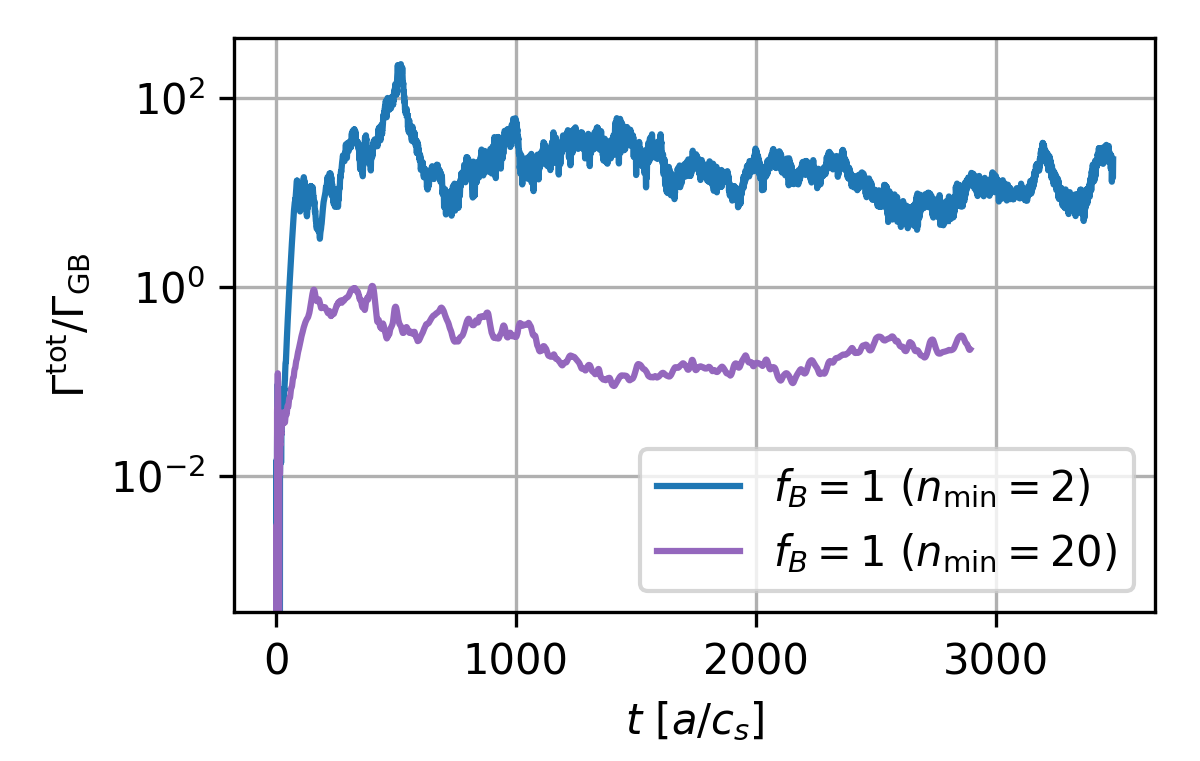}}
        \captionsetup{font=it}
    \caption{Time traces of the total heat flux \emph{(\textbf{a})} and total particle flux \emph{(\textbf{b})} from \texttt{GENE} simulations. Both simulations include $\delta B_\parallel$ and use the exact prescription of the magnetic-drift velocity. The value $n_{\mathrm{min}}$ is the toroidal mode number corresponding to the value of $k_{y,\mathrm{min}}.$ The simulation with the larger value of $n_{\mathrm{min}}$ only evolves those unstable modes that are also unstable when the MHD-1 approximation is used.  Removing the lowest $k_y$ modes (purple) in this way results in saturation at values two orders of magnitude lower than the baseline (blue).}
    \label{fig:kymin_comp}
\end{figure}

Fig~\ref{fig:bpar_comp_sec5} shows time traces of the total heat flux from two nonlinear \texttt{GENE} simulations. The first simulation (blue) includes $\dbp$ and uses (\ref{eqn:full_drift}) for the drift velocity (this simulation is equivalent to that shown in Fig 3a of \cite{giacomin2023b} and will henceforth be referenced as the `STEP baseline' simulation). The second simulation (green) shown in Fig~\ref{fig:bpar_comp_sec5} does not include $\dbp$ and instead uses the MHD approximation (MHD-1). Although the linear physics is broadly similar (both linear spectra have hKBMs as the dominant unstable mode), the nonlinear physics produced by these two simulations is strikingly different. The simulation using the MHD approximation appears to achieve a robustly steady\footnote[1]{In this paper, we use the same definition of robustly steady as was introduced in \cite{giacomin2023b}. The augmented Dickey-Fuller statistical test~\cite{dickey1979} is applied to the time trace of the total heat flux in the saturated or pseudo-saturated phase of each simulation (i.e. after the linear growth phase) to determine whether the saturation corresponds to a robust stationary state. The null hypothesis of the augmented Dickey-Fuller test is the presence of a unit root, while the alternative hypothesis is the stationarity of the time series. The null hypothesis is rejected when the \texttt{p-value} returned by the statistical test is below a threshold value that is taken to be \texttt{p}$ = 0.1$ in this paper. {The only simulation in this paper that does not saturate by this criterion is the reduced $\beta^{\prime}$ calculation shown in Figure~\ref{fig:kymin_comp}.}} saturated state at values of the heat flux around two orders of magnitude smaller than those reached by the simulation that includes full $\dbp$ physics. {It is interesting to note that the lower flux MHD approximation simulation does not exhibit the high-frequency oscillations, seen in the full $\dbp$ physics case, that appear to be typical of such simulations (see \cite{giacomin2023b}); this is under investigation}. Fig~\ref{fig:bpar_comp_sec5} also shows the time trace of the $k_y$ spectrum of $\delta \! \phi$ and $\delta \! A_\parallel$ for the two simulations. In both cases, the zonal mode dominates over the non-zonal modes and it is likely that zonal flows and fields play a role in saturating the hKBM instability\footnote{The exact saturation mechanism of the hKBM in the absence of equilibrium flow shear demands further study which will be the focus of future work.}. Despite the difference between the two simulations, it is important to remark that the heat flux predicted using the MHD approximation is still orders of magnitude larger than the heat flux driven by the subdominant MTM (see Fig 14 of \cite{giacomin2023b}), showing the strong effect of the hKBM instability when the MHD approximation is used to drop $\dbp$. Snapshots of the turbulence are shown in \ref{app:contours}, Fig~\ref{fig:turbulence_snapshots}. 
{The considerably reduced transport fluxes with the MHD approximation, compared to with full physics, are associated with lower amplitude turbulent fluctuations shown in Fig.\ref{fig:bpar_comp_sec5}, and less radially extended turbulent structures in the perturbed fields (see Fig.\ref{fig:turbulence_snapshots}(a-f)).} 

The reason for saturation at much lower fluxes under the MHD approximation in Fig~\ref{fig:bpar_comp_sec5} is revealed by closer inspection of the MHD-1 linear results in Fig~\ref{fig:bpar_comp2}a. From this figure, we can see that long wavelength modes, which are unstable with $\dbp$, are stable with the MHD approximation, and it is precisely these wavenumbers at $k_y \rho_D \ll 1,$ that are associated with the very large turbulent fluxes. To be explicit, it appears to be possible to find a lower-flux saturated state even when hKBMs are unstable across a wide range of the linear spectrum provided that they are not unstable up to very long wavelengths. Further evidence for this claim is demonstrated in Fig~\ref{fig:kymin_comp}. This figure compares the total heat and particle fluxes from the baseline simulation (blue) to a second simulation (purple) that also includes the full-physics model but on a reduced grid of poloidal wavenumbers such that the only unstable modes evolved are those which are also unstable when the MHD-1 approximation is used. Removing the low-$k_y$ modes results in saturation at much smaller levels even when all of the retained modes have identical linear properties.

The results in Fig~\ref{fig:bpar_comp_sec5} appear to support the conclusions of \cite{kennedy2023a,giacomin2023b} that $\dbp$ fluctuations play an essential role in hKBM driven turbulence in these STEP local equilibria, and in particular in the transition to a large-transport state. However, it has been shown \cite{kennedy2023a,giacomin2023b} that it is possible to vary the local equilibrium in such as way as to change the hKBM threshold. In Section~\ref{sec:parallel} we study whether it is possible to extend this result by posing the question;  can we find local equilibria where the hKBM is linearly unstable in the absence of $\dbp$ with exact drifts? 

\section{The importance of parallel magnetic perturbations to the hKBM}
\label{sec:parallel}

In \cite{giacomin2023b} it was shown that the total heat and particle fluxes from nonlinear simulations vary with $\beta$ when $\beta^\prime$ is varied consistently (see Fig. 11 of \cite{giacomin2023b}). However, the heat and particle fluxes remained very large even at much smaller values of $\beta$ than the STEP baseline. It was argued that reduced $\beta^{\prime}$ stabilisation of the hKBM is largely responsible for such large turbulent fluxes at lower $\beta$. In the STEP baseline case, $\beta^\prime$ stabilisation is very important since it pushes the hKBM towards marginality. The sensitive dependence of the hKBM on $\beta^\prime$
is made most obvious when $\beta^\prime$ is varied holding all other local equilibrium parameters fixed (albeit inconsistent). {It is interesting to note, as remarked in Section~\ref{sec:betaprime_discussion}, that we should expect the MHD approximation to improve as $|\beta^\prime|$ is reduced.} 

\begin{figure}[htbp]
    \centering
    \subfloat[]{\includegraphics[width=0.48\textwidth]{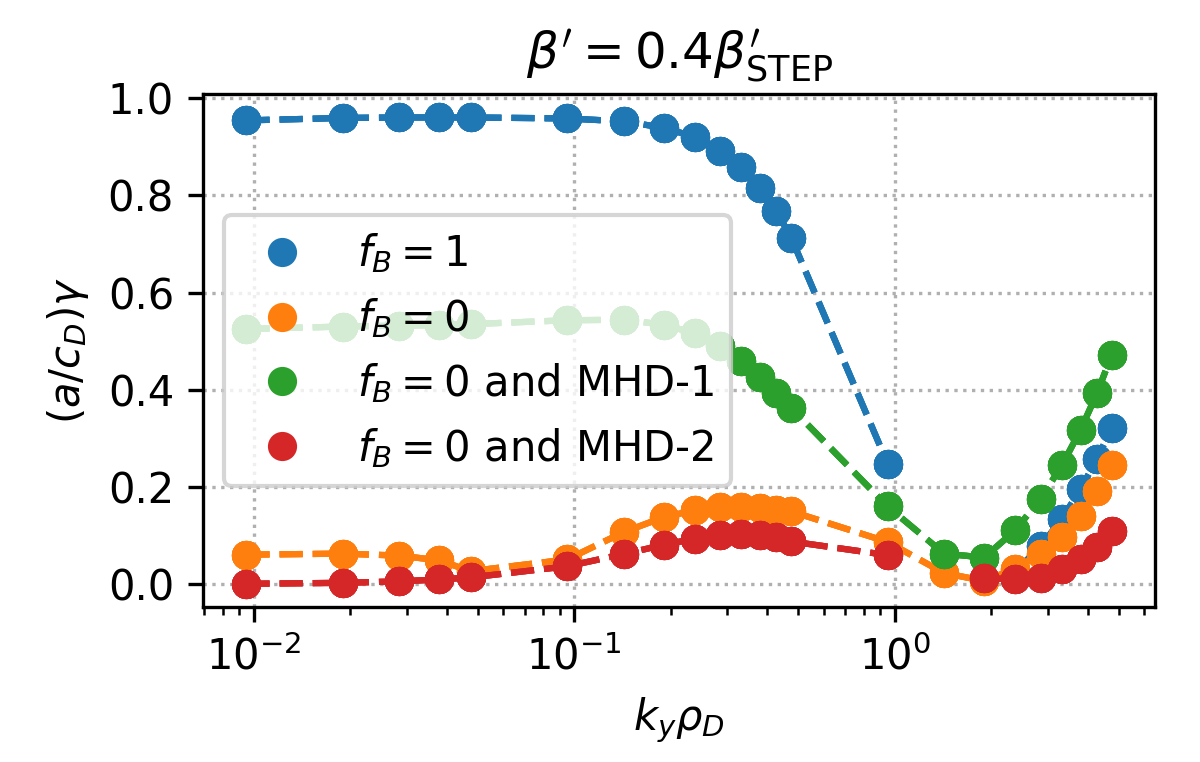}}\quad
   \subfloat[]{\includegraphics[width=0.48\textwidth]{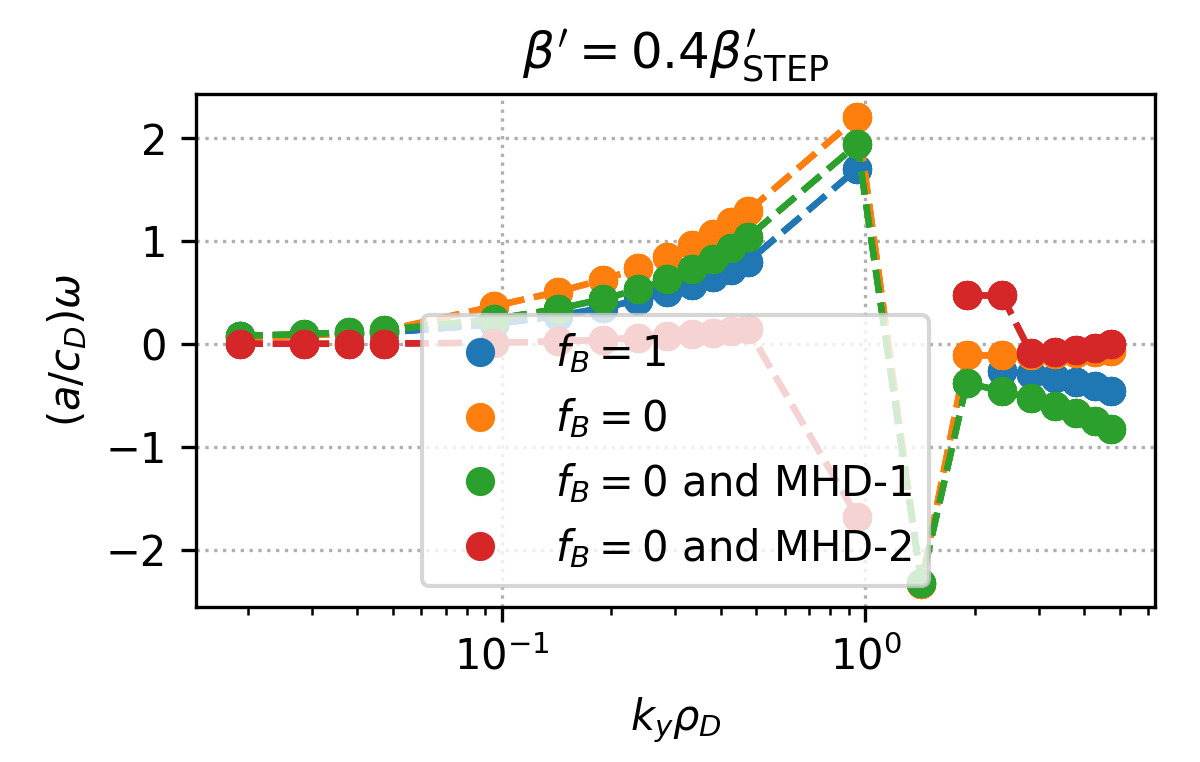}}\\
    \subfloat[]{\includegraphics[width=0.48\textwidth]{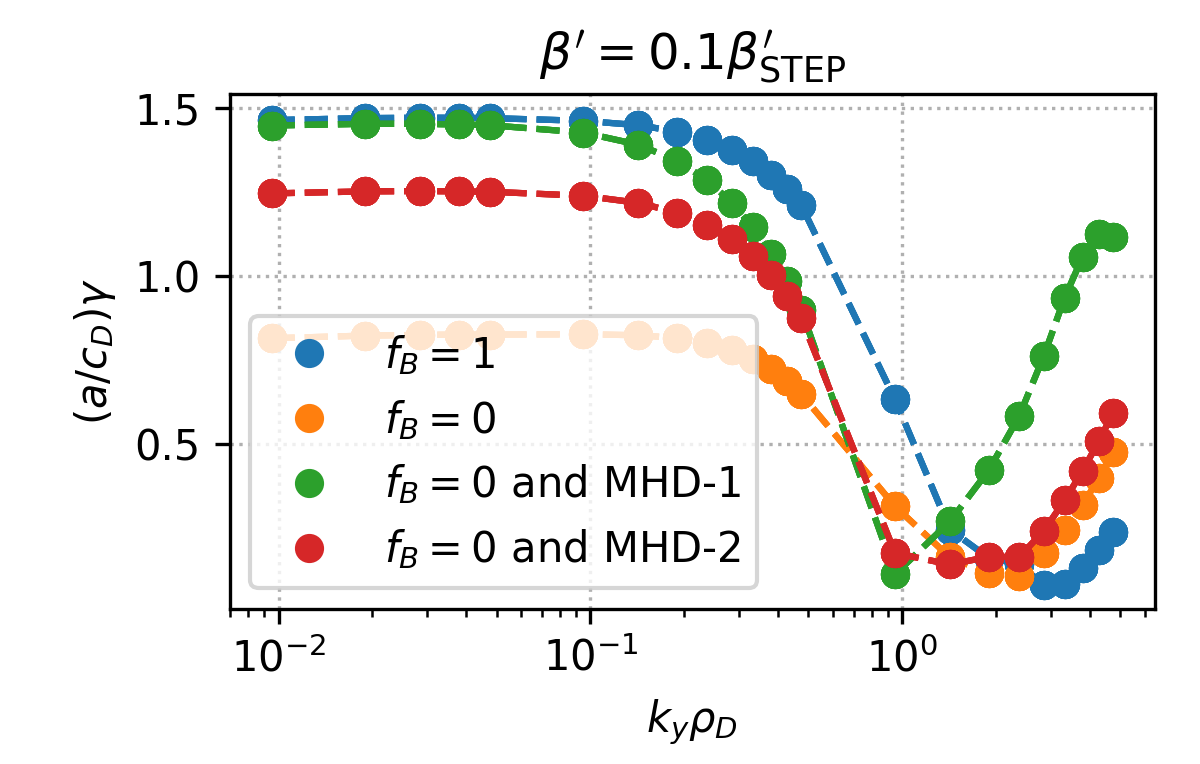}}\quad
    \subfloat[]{\includegraphics[width=0.48\textwidth]{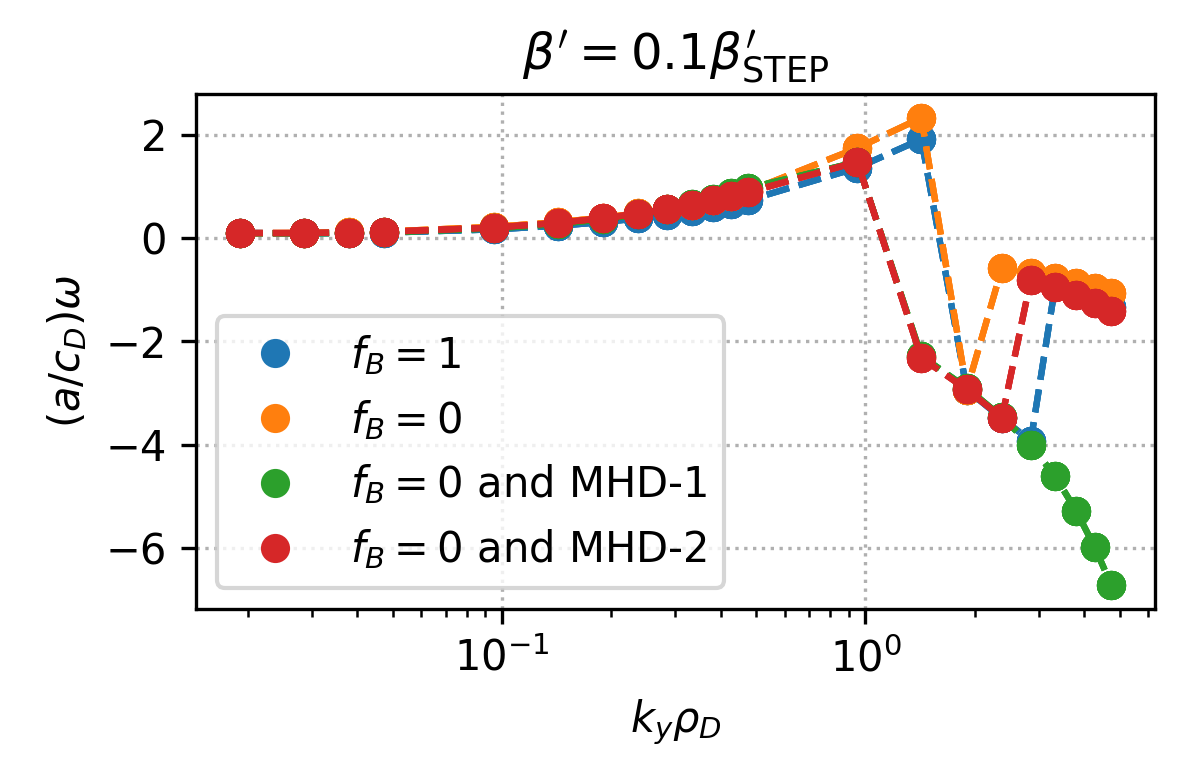}}
    \captionsetup{font=it}
     \caption{Growth rates \emph{(\textbf{a, c})} and mode frequencies \emph{(\textbf{b, d})} as functions of the binormal wavenumber from linear simulations of the dominant instability in STEP-EC-HD on a mid-radius flux surface. Results are shown with reduced $\beta^\prime = 0.4 \beta^\prime_{\mathrm{STEP}}$ \emph{(\textbf{a, b})} and $\beta^\prime = 0.1 \beta^\prime_{\mathrm{STEP}}$ \emph{(\textbf{c, d})}. All other parameters are kept fixed.  Simulations are shown both with, $f_B=1$ (blue), and without, $f_B=0$ (orange, green, red), $\delta \! B_\parallel$ fluctuations. For the simulations without $\dbp,$ different treatments of the drift velocity are also shown: (i) MHD-1 (green); and (ii) MHD-2 (red). In all cases, the hKBM is much more unstable than the reference case (see Fig~\ref{fig:bpar_comp2}) due to reduced $\beta^\prime$ stabilisation \cite{bourdelle2005}.}
    \label{fig:reduced_betaprime}
\end{figure}

Fig~\ref{fig:reduced_betaprime} shows the linear growth rate ({a, c}) and frequency ({b, d}) as functions of
the binormal wavenumber for simulations with reduced $\beta^\prime$ with, $f_B=1$ (blue), and without, $f_B=0$ (orange, green, red), $\dbp$ fluctuations. In \textbf{all} cases the hKBM is much more unstable than at the nominal value of $\beta^\prime$ (see Fig.~\ref{fig:bpar_comp2}) due to the much decreased role of $\beta^\prime$ stabilisation. In each case, the hKBM is likely to drive much more transport.

There are several key observations from Fig~\ref{fig:reduced_betaprime} that we wish to stress.

\begin{itemize}
\item In the limit $\beta^\prime \rightarrow 0,$ the MHD-1 and MHD-2 are identical, which is not surprising as $\bb \cdot \mathbf{\gradd} \bb \rightarrow (\mathbf{\gradd_{\perp}} \B)/\B$ as $\mathbf{\gradd} p \rightarrow 0$ (see \ref{sec:appendixc}). This is clearly seen in Fig~\ref{fig:reduced_betaprime} where the difference between the MHD-1 and MHD-2 prediction reduces as $|\beta^\prime|$ decreases. {The limit $|\beta^\prime| \rightarrow 0$ is in effect a $\omega_{\gradd B} - \omega_{\kappa} \rightarrow 0$ approximation. In this physical limit $\dbp=0$, and the MHD approximation should be trivially satisfied.}

    \item We note that the MHD approximation does a much better job of capturing the linear growth rates of the hKBM at smaller values of $|\beta^\prime|$, {in agreement with the discussion in Section~\ref{sec:betaprime_discussion}.} 


\item  The hKBM can be unstable up to $n = 1$ in the absence of $\dbp$ with exact drifts, and this is certainly a no-go zone for any actual device. 
\end{itemize}
In more strongly driven regimes at lower $\beta^{\prime}$ in STEP, it is clearly possible to study hKBM-driven turbulence using global GK codes that do not include $B_{\parallel}$.

\begin{figure}
\centering
   \subfloat[]{\includegraphics[width=0.48\textwidth]{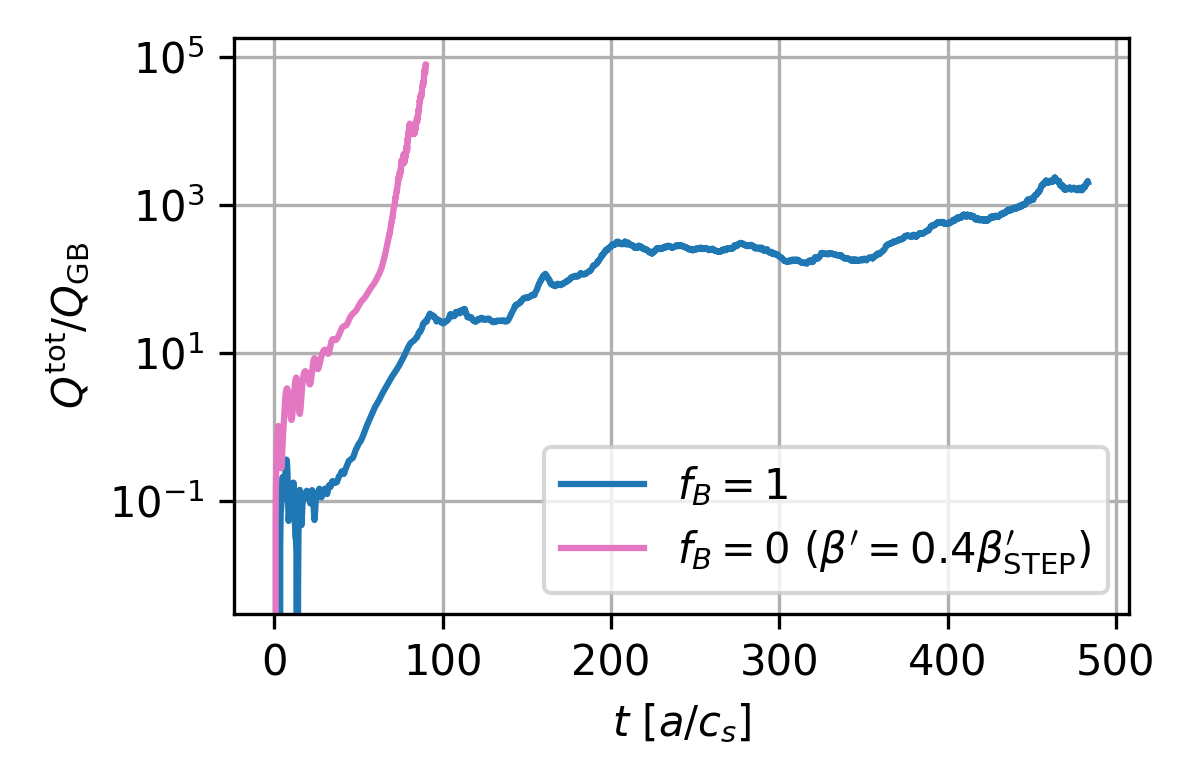}}\quad
    \subfloat[]{\includegraphics[width=0.48\textwidth]{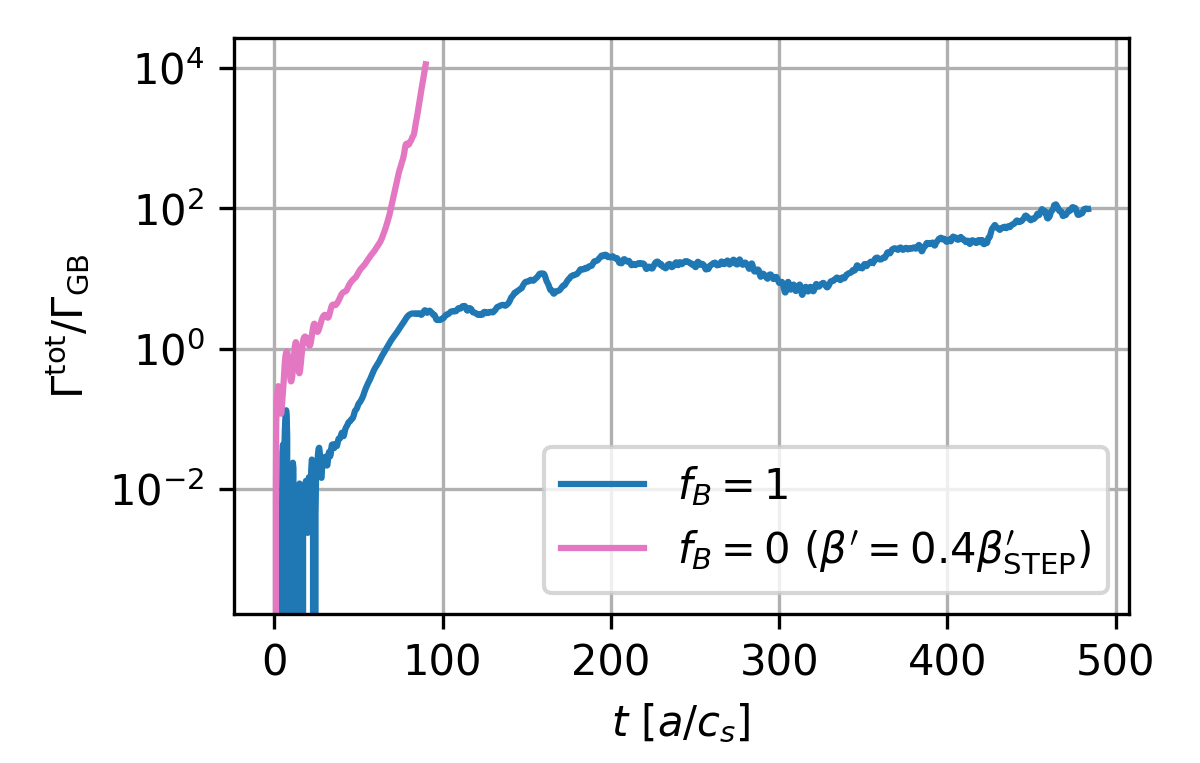}}
        \captionsetup{font=it}
    \caption{Time traces of the total heat flux \emph{(\textbf{a})} and total particle flux \emph{(\textbf{b})} from \texttt{GENE} simulations. Simulations are shown both with $\delta B_\parallel$ at the nominal $\beta^{\prime}$ ($f_B=1$, blue), and without at $\beta^\prime = 0.4 \beta^\prime_{\mathrm{STEP}}$ ($f_B=0$, magenta). No approximations are made concerning the magnetic-drift velocity. {The lower $\beta^{\prime}$ simulation without $\dbp$ (magenta) exhibits a qualitatively similar electromagnetic runaway (transition to very large heat fluxes) to that seen with full physics since the hKBM has been driven unstable at long wavelength, but it is the only simulation in this paper that does not meet the statistical definition of saturation introduced in \cite{giacomin2023b}.}}
    \label{fig:bpar_comp3}
\end{figure} 

Fig~\ref{fig:bpar_comp3} shows time traces of the total heat flux and particle flux from two nonlinear \texttt{GENE} simulations: (i) with the nominal value of $\beta^\prime$ that includes $\dbp$ fluctuations (blue); and (ii) excluding $\dbp$ fluctuations and using exact drifts (magenta) at a lower value of $\beta^\prime = 0.4 \beta^\prime_{\mathrm{STEP}}$ (with all other parameters are identical). As expected from the linear spectrum (Fig~\ref{fig:reduced_betaprime}), the nonlinear physics of this new case qualitatively mirrors the STEP baseline case even without $\dbp$ fluctuations. The heat fluxes rise to very large levels and no robustly steady saturated state is reached over the period of this simulation: (note that this is the only simulation in this paper that does not reach a robustly steady saturated state). {It is currently unknown whether this simulation would obtain a steady state if it were continued and such a question is very difficult to answer numerically due to the computational cost. Simulations such as those discussed here motivate the pressing need for a rigorous theory of electromagnetic turbulence saturation in the flux-tube limit.} Snapshots of the turbulent fields Fig~\ref{fig:turbulence_snapshots}(g,h,i) reveal the presence of box-scale radially elongated streamers. 

{The fact that at lower $\beta^{\prime}$ we obtain electromagnetic runaway fluxes without $\dbp$ is an important result; it reduces the number of ingredients needed to observe electromagnetic runaway fluxes in local gyrokinetic simulations. Put more simply, we observe qualitatively similar runaway fluxes in local simulations when there is electromagnetic instability at the longest-wavelengths in the system. We take great care to emphasise that the results in this paper suggests that there is an important distinction to be made between (A) runaway fluxes that eventually saturate at very large values such as those discussed in this paper, and (B) fluxes that never saturate such as the runaways reported by \cite{pueschel2013} for cyclone-base-case\cite{cbc} geometry without $\dbp.$ All nonlinear results for the STEP equilibrium at the nominal $\beta^{\prime}$ in this paper definitively fall into category (A). It remains unclear which category the lower $\beta^\prime$ simulation falls into. Identification of these two categories motivates the need for a rigorous theory of electromagnetic turbulence saturation that can distinguish between the two types of behaviour.}   

\section{Conclusions}
\label{sec:conclusions}
In this paper, we have examined the importance of including parallel magnetic perturbations ($\dbp$) in gyrokinetic simulations of electromagnetic turbulence at mid-radius in the burning plasma phase of the conceptual high-$\beta$ reactor-scale, tight-aspect-ratio tokamak STEP. It has previously been found~\cite{kennedy2023a} that $\dbp$ is essential for the hKBM to be unstable in local equilibria taken from the STEP burning flat top. 

\subsection{The MHD approximation is typically not appropriate for use in high-\texorpdfstring{$\beta_e$}{TEXT} reactor-scale plasmas}

{In Section~\ref{sec:betaprime_discussion} it was argued that the MHD approximation\footnote{The MHD approximation is faithfully represented by the implementation MHD-1, discussed in the main text, which compensates for the neglect of parallel magnetic perturbations by setting $\omega_{\gradd B} = \omega_{\kappa}$.} should perform well in the limit $\beta^\prime \ll \sqrt{\beta} \ll 1$.  This limit is not well satisfied in STEP-EC-HD (where $\beta_e=0.09$ and $|\beta^{\prime}|=0.48$ at $\rho/a=0.64$), and it may also be questionable in existing experiments at internal/external transport barriers with high pressure gradients, and in other reactor-scale plasmas.}

In Section~\ref{sec:linear_GK}, it was shown that inclusion of $\dbp$ and a rigorous treatment of the magnetic drift velocity is essential to capture the correct linear spectrum of the hKBM in the STEP flat-top. The MHD approximation roughly captures the linear spectrum of the hKBM, but predicts lower growth rates  and is significantly more stable at long wavelengths for this STEP local equilibrium. 

Even though the MHD approximation reproduces qualitatively similar linear physics, Section~\ref{sec:nonlinear_gk}  reveals that the corresponding nonlinear simulations do not accurately track nonlinear simulations for STEP with full physics that were reported in \cite{giacomin2023b}. Typically, simulations which attempt to resolve the hKBM with full physics (including $\dbp$) yield fluxes rising
to very large values and require substantial computational resource to reach saturation.
When the MHD approximation is used, the turbulence saturates at a level around two orders of magnitude lower than that reached in a comparable time in the full physics simulations. The reason for this difference is that the longest wavelength modes (i.e., corresponding to $n < 10$) are stable under the MHD approximation, but unstable with full physics. It was argued that these linearly unstable modes at very long-wavelength are responsible for the transition to very large fluxes (see Fig~\ref{fig:kymin_comp} and discussion).

\subsection{\texorpdfstring{$\dbp$}{TEXT} is not always essential for studying hKBM driven turbulence. However, \texorpdfstring{$\dbp$}{TEXT} is essential for studying turbulence in the high $\beta^{\prime}$ conditions of the STEP flat-top.}

In Section~\ref{sec:parallel}, strong $\beta^\prime$ stabilisation \cite{bourdelle2005} was identified as being responsible for the sensitivity of hKBM stability to $\dbp$ fluctuations, and it was noted that this plasma sits close to marginality. It was also shown, however, that the hybrid mode becomes unstable in the absence of $\dbp$ if $\beta^{\prime}$ is reduced to push the local equilibrium further from marginality. In such local equilibria, which are more unstable at long wavelength, electromagnetic runaway fluxes emerge in nonlinear simulations that neglect $\dbp$ (without, and likely also with, the MHD approximation). This suggest that turbulence can be simulated in some high-$\beta$ ST plasmas using presently available tools and codes that neglect $\delta \!B_{\parallel}$.  

In the present design of the STEP flat-top, however, it is essential to include $\delta \! B_{\parallel}$ because the modes expected to dominate turbulence are both: (i) unstable only with $\dbp$ and; (ii) unstable up to very long wavelengths. Global gyrokinetic codes currently being developed to include $\dbp$ rigorously will be important and timely for STEP.

Finally this work motivates the need for a rigorous theory of electromagnetic turbulence saturation, and an elucidation of the role of $\dap$ in particular. 

\ack

Simulations were performed using resources provided by the Cambridge Service for Data Driven Discovery (CSD3) operated by the University of Cambridge Research Computing Service (\url{www.csd3.cam.ac.uk}), provided by Dell EMC and Intel using Tier-2 funding from the Engineering and Physical Sciences Research Council (capital grant EP/T022159/1), and DiRAC funding from the Science and Technology Facilities Council (\url{www.dirac.ac.uk}). The work of PI was supported by the Engineering and Physical Sciences Research Council (EPSRC) [EP/R034737/1]. The work of TA was supported in part by the Royal Society Te Apārangi Marsden grant MFP-UOO2221.  The authors (CMR) would also like to acknowledge helpful discussions with R J Hastie and J W Connor.

To obtain further information on the data and models underlying this paper please contact PublicationsManager@ukaea.uk.

\appendix

\section{The MHD approximation in the GK framework}
\label{sec:appendixA}
The perturbed perpendicular pressure $\delta \! P_{\perp}$ appearing on the right hand side of the linear GK force balance equation~\ref{eqn:pressure_balance} can be calculated by solving the linear gyrokinetic equation~\ref{eqn:linear_gk} in the limit $k_{\parallel}v_{\mathrm{th}s} \ll \omega_{ds} \ll \omega$ and substituting into equation~\ref{eqn:dbp} to give
\begin{eqnarray}
\delta \! P_{\perp} = \sum_{s} \int \dv \, \frac{m_{s}q_{s}\vpp^{2}}{2T_{0s}} &\left[ 1 - \frac{\omega_{\star s}^{T}}{\omega} + \frac{\omega_{ds}}{\omega} \right] \nonumber \\
   &\times  \left[ \delta\!\phi - v_{\parallel}\delta \! A_{\parallel} + \frac{m_{s}v_{\perp}^{2}}{2q_{s}\B} \dbp \right] F_{0s}.
\end{eqnarray}
Replacing $\dbp$ with $\delta \! P_\perp$ via (\ref{eqn:pressure_balance}), and dropping terms involving $\delta\!A_{\parallel}$ that vanish because they are odd in $v_{\parallel}$, we can write:
\begin{eqnarray}
     \delta P_{\perp} = \sum_{s} \int \dv \,  F_{0s} \frac{m_{s} \vpp^{2} q_{s}}{2T_{0s}} \left[1 - \frac{\omega_{\star s}^{T}}{\omega} + \frac{\omega_{ds}}{\omega}\right] \times \left[ \delta\!\phi - \frac{m_{s}v_{\perp}^{2}}{2q_{s}} \frac{\mu_{0} \delta P_{\perp}}{\B^2}\right].
\end{eqnarray}
That is,  we obtain
{\setlength{\mathindent}{0cm} \small
\begin{eqnarray}
    \delta \! P_{\perp} &=& \sum_s \int \dv \, F_{0s} \left\{ \frac{m_{s} \vpp^{2}}{2T_{0s}}q_{s} \delta \! \phi\left[ 1 -\frac{\omega_{\star s}^{T}}{\omega} + \frac{\omega_{ds}}{\omega} \right] \right. - \left. \frac{m_s^2 \vpp^4}{4 T_{0s}} \frac{\mu_0 \delta P_\perp}{\B^2} \left[ 1 -\frac{\omega_{\star s}^{T}}{\omega} + \frac{\omega_{ds}}{\omega} \right] \right\} \label{eqn:dpperp} \\
    &=& \underbrace{\delta \! \phi  \sum_s \int \dv \, \frac{q_{s}  m_{s} \vpp^{2} F_{0s}}{2T_{0s}} \left[ 1 -\frac{\omega_{\star s}^{T}}{\omega} + \frac{\omega_{ds}}{\omega} \right]}_{[I]} - \underbrace{\frac{\mu_0 \delta \!P_\perp}{\B^2} \sum_s \int \dv \, \frac{m_s^2 \vpp^4 F_{0s}}{4 T_{0s}}  \left[ 1 -\frac{\omega_{\star s}^{T}}{\omega} + \frac{\omega_{ds}}{\omega} \right]}_{[II]} \nonumber
\end{eqnarray}
Performing the velocity space integrals and sums over species allows us to obtain:
\begin{eqnarray}
    [I] & = &  \delta \! \phi \sum_s n_s q_s \left\{ 1 -\frac{\omega_{\star s}}{\omega}\left( 1 + \eta_s \right) + \frac{T_{0s}}{q_s \B \omega} \left( \omega_{\kappa}+2 \omega_{\nabla B} \right) \right\}. \label{eq:I1} 
\end{eqnarray}
The first term in (\ref{eq:I1}) vanishes due to quasineutrality. On summing over species the numerator in the second term reduces to:
\begin{equation}
    \sum_s n_s q_s \omega_{\star s}(1+\eta_s)= \frac{(\mathbf{k}\times\bb \cdot \gradd{P})}{\B} = \frac{\B}{\mu_0} \left( \omega_{\kappa}-\omega_{\nabla B}\right),
\end{equation}
where we have also used (\ref{eq:omkap-omgb}) relating $\omega_{\kappa}$ and $\omega_{\nabla B}$, derived in \ref{sec:appendixc}.  Using these results in (\ref{eq:I1}) we obtain:
\begin{eqnarray}
    [I] & = \frac{\B \delta \! \phi}{\mu_0 \omega} \left( \left(\omega_{\nabla B} - \omega_{\kappa}\right) + \frac{\beta}{2} 
    \left( \omega_{\kappa}+2 \omega_{\nabla B} \right) \right) \label{eq:I} 
\end{eqnarray}
Evaluating similar velocity space integrals in $[II]$ from the RHS of \ref{eqn:dpperp}, gives:
\begin{equation}
    [II] = - \delta P_{\perp} \sum_s \beta_s \left[ 1 - \frac{\omega_{\star s}}{\omega} (1+2 \eta_s) + \frac{T_{0s}}{q_s \B \omega} (\omega_{\kappa} + 3 \omega_{\nabla B}) \right] \label{eq:II}
\end{equation}
Now substituting for $[I]$ and $[II]$ [from (\ref{eq:I}) and (\ref{eq:II})] into (\ref{eqn:dpperp}), retaining only leading order terms in a small $\beta$ expansion gives: 
\begin{equation}
    \delta P_{\perp} = \frac{\B \delta \! \phi}{\mu_0 \omega} \left(\omega_{\nabla B} - \omega_{\kappa}\right),
    \label{eq:A8}
\end{equation}
obtained in the limits $k_{\parallel} v_{\mathrm{th}s} \ll \omega_{ds} \ll \omega$ and $\beta \ll 1$. It follows from (\ref{eq:A8}) that in these limits $\delta \!P_{\perp}$ and therefore $\dbp$, can only be neglected if: 
\begin{equation}
    \omega_{\nabla B} = \omega_{\kappa}
\end{equation}
This is the MHD approximation that appears as (\ref{eq:MHD_approx}) in the main text.

\section{Cancellation of Magnetic Drift and $\mathbf{\delta \! B_{\parallel}}$ Terms in Appropriate Limits}
\label{sec:appendixb}
Using the relationship between $\gradd B$ and curvature drifts in \ref{eq:omkap-omgb}, the drift term on the left-hand side of the GK equation in \ref{eqn:linear_gk} can be written in a form convenient for the MHD-1 approximation:
\begin{eqnarray}
    T_{ds} & = & -\frac{1}{\Omega_{s}} \left[\omega_{\kappa} \left( v_{\parallel}^{2} + \frac{v_{\perp}^{2}}{2} \right) - \frac{\mu_0 (\mathbf{k}_\perp \cdot \bb \times \gradd p)}{\B^2}\frac{v_{\perp}^2}{2}\right] g_s 
    \label{B1}
\end{eqnarray}
The final term is the pressure gradient contribution to the $\gradd B$ drift:
\begin{eqnarray}
    T_{\nabla B}^{p^{\prime}} & = & \frac{1}{q_{s} \B} \frac{\mu_0 \mathbf{k}_\perp \cdot \bb \times \gradd p}{\B^2}\frac{m_sv_{\perp}^2}{2} g_s  =  \omega_{\rm dia} \frac{\mu_0 n_{0s} T_{0s}}{\B^2} \frac{m_s v_{\perp}^2}{2 T_{0s}} g_s \label{eq:GBpprime}
\end{eqnarray}
where $\omega_{\rm dia}=\mathbf{k}_\perp \cdot \mathbf{V_{dia}}$ and the diamagnetic velocity $\mathbf{V_{dia}}=\bb\times \mathbf{\gradd} p/(n_{0s} q_s \B)$.

Alternatively the magnetic drift term can be expressed as:
\begin{eqnarray}
    T_{ds} & = & -\frac{1}{\Omega_{s}} \left[\omega_{\gradd B} \left( v_{\parallel}^{2} + \frac{v_{\perp}^{2}}{2} \right) + \frac{\mu_0 (\mathbf{k}_\perp \cdot \bb \times \gradd p)}{\B^2}v_{\parallel}^2\right] g_s . 
    \label{B3}
\end{eqnarray}
where the final term resembles a pressure gradient contribution to the curvature drift (though this decomposition is unphysical) and can be written as:
\begin{eqnarray}
    T_{\kappa}^{p^{\prime}} & = & -\omega_{dia} \frac{\mu_0 n_{0s} T_{0s}}{\B^2} \frac{m_s v_{\parallel}^2}{T_{0s}} g_s
\end{eqnarray}

In the long wavelength high frequency limit ($b_s \rightarrow 0$ and $\omega \gg \omega_{\star s}^T$) the compressional magnetic perturbation term on the RHS of the GK equation reduces to:
\begin{eqnarray}
T_{\delta \! B_\parallel}  & = & \omega \frac{\delta \! B_{\parallel}}{\B} \frac{m_{s}v_{\perp}^{2}}{2 T_{0s} } F_{0s} = - \omega \frac{\mu_0 \delta \!P_{\perp}}{\B^2} \frac{m_{s}v_{\perp}^{2}}{2 T_{0s} } F_{0s},
\end{eqnarray}
using perturbed perpendicular force balance in equation~\ref{eqn:pressure_balance} to reach the second equality.

\vspace*{0.2cm}
\noindent {\em Consideration of MHD-1}: In the long wavelength fluid limit,  $g_s$ is Maxwellian  to leading order in $\omega_{ds}/\omega$, so $T_{\gradd B}^{p^\prime}$ and $T_{\delta \! B_\parallel}$ have approximately the same structure in velocity space.  A cancellation between $T_{\nabla B}^{p^\prime}$ and $T_{\delta \! B_\parallel}$ will arise at leading order if:
\begin{equation}
    \frac{q_s \delta \! \phi}{T_{0s}}= -\frac{\omega}{\omega_{dia}} \frac{\delta P_{\perp}}{n_s T_{0s}},
    \label{eqn:B6}
\end{equation}
This approximate cancellation between $T_{\nabla B}^{p^\prime}$ and $T_{\delta \! B_\parallel}$ allows the $\gradd p$ contribution to the $\gradd B$ drift and $\delta \! B_{\parallel}$ to be dropped from the GK equation, and corresponds to MHD-1.

\vspace*{0.2cm}
\noindent{\em Consideration of MHD-2:} On the other hand,  $T_{\kappa}^{\gradd p}$ and $T_{\delta \! B_\parallel}$ have manifestly different structure in velocity space, so exact cancellation is not possible in the GK equation.  While MHD-2 implements \ref{eq:MHD_approx} to allow $\delta\!B_{\parallel}$ to be dropped from the GK equation, it only does this at the expense of {\em changing the local equilibrium and its magnetic curvature, $\bb \cdot \nabla \bb$}. In effect MHD-2 {\em changes the local equilibrium field curvature}, and {\em again removes the pressure gradient contribution to the $\nabla B$ drift}, $T_{\nabla B}^{p^{\prime}}$ defined in \ref{eq:GBpprime},  for this modified local equilibrium!

\section{A useful identity involving the magnetic-drift frequency}
\label{sec:appendixc}

The magnetic-drift frequency is given by:
{\setlength{\mathindent}{0.5cm} \small
 \begin{eqnarray}
     \omega_{ds} 
     &= \mathbf{k}_\perp \cdot \bb \times \left( \frac{v_{\parallel}^{2}}{\Omega_s} (\bb\cdot\gradd)\bb + \frac{\vpp^2 }{2\Omega_s} \nabla \ln B \right) =  \frac{\omega_{\kappa} v_{\parallel}^2}{\Omega_s}  + \frac{\omega_{\nabla B} v_{\perp}^2}{2 \Omega_s} \label{eqn:full_drift_b1}
 \end{eqnarray}}
Using Amp{\`e}re's law for the equilibrium, we can write 
 \begin{eqnarray}
     \mu_0 \mathbf{J}\times\mathbf{B} &= [\gradd\times\mathbf{B}] \times \mathbf{B} \nonumber \\ &= -\mathbf{B} \times [\gradd\times(\B\bb)] \nonumber \\ 
     &= -\mathbf{B} \times [\gradd \B \times \bb + \B(\gradd\times\bb)] \nonumber \\ 
     &= -\mathbf{B} \times [\gradd \B \times \bb] -\mathbf{B} \times \B (\gradd \times \bb)] \nonumber \\ 
     &= - \B\gradd \B + (\mathbf{B} \cdot \gradd \B)\bb - \B^2 \bb \times (\gradd \times \bb) \nonumber \\
   \Rightarrow \mu_0 \mathbf{J}\times\mathbf{B}  &= - \B\gradd_{\perp} \B + \B^2 \bb \cdot \gradd \bb \label{eq:JxB}
 \end{eqnarray}
 where $\bb$ is a unit vector aligned to the equilibrium magnetic field, so that $\bb \times (\gradd \times \bb)= -\bb \cdot \gradd \bb$.  Now invoking equilibrium force balance $ \mathbf{J}\times\mathbf{B}=\gradd p$ in (\ref{eq:JxB}) gives: 
\begin{eqnarray}
    \gradd_{\perp} \ln B &= (\bb\cdot \gradd)\bb - \frac{\mu_0 \gradd p}{\B^2}.  \label{eq:gbcurv}
\end{eqnarray}
which usefully relates the gradient in magnetic-field strength to the curvature, and can be expressed as:
\begin{equation}
    \omega_{\kappa} -\omega_{\nabla B} = \frac{\mu_0 \mathbf{k}_\perp \cdot \bb \times \gradd p}{\B^2} \label{eq:omkap-omgb}
\label{C4} \end{equation}

Using (\ref{eq:gbcurv}) to substitute for $\gradd \ln B$ in (\ref{eqn:full_drift_b1}), the magnetic drift frequency can be re-expressed as: 
\begin{eqnarray}
    \Omega_s \omega_{ds} &= \mathbf{k}_\perp \cdot \bb \times \left[ \left( v_{\parallel}^{2} + \frac{\vpp^2 }{2} \right) (\bb\cdot\gradd)\bb - \frac{\vpp^2 \mu_0 \gradd p}{2\B^2}  \right]  .  
\end{eqnarray}

\newpage
\section{Snapshots of the turbulent fields}
\label{app:contours}

    \begin{figure}[htbp]
    \centering
    
      \subfloat[$f_{B}$ = 1]{\includegraphics[width=0.3\textwidth]{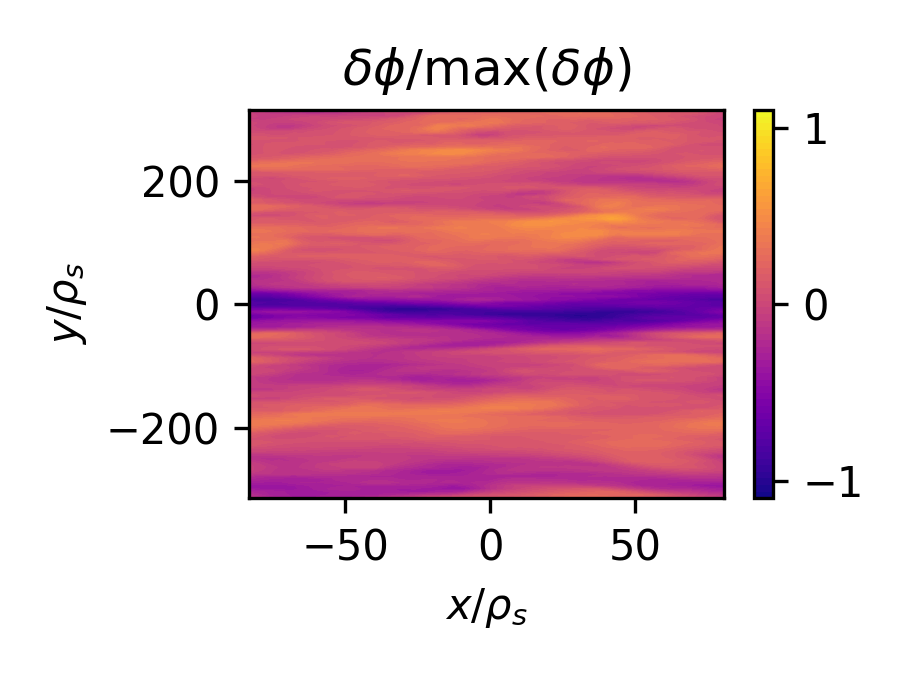}}\quad
   \subfloat[$f_B = 1$]{\includegraphics[width=0.3\textwidth]{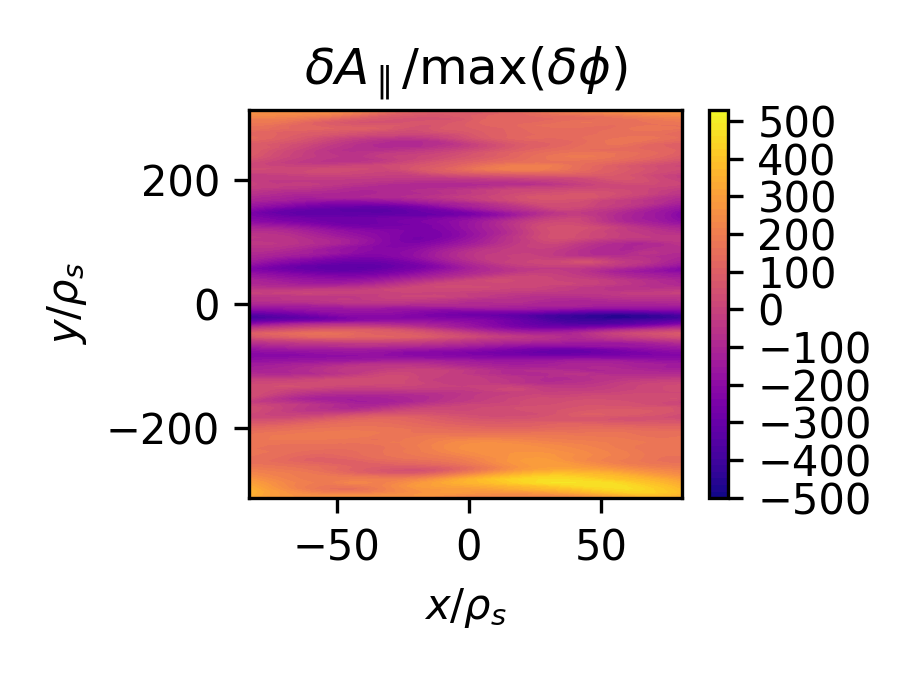}} \quad \subfloat[$f_B = 1$]{\includegraphics[width=0.3\textwidth]{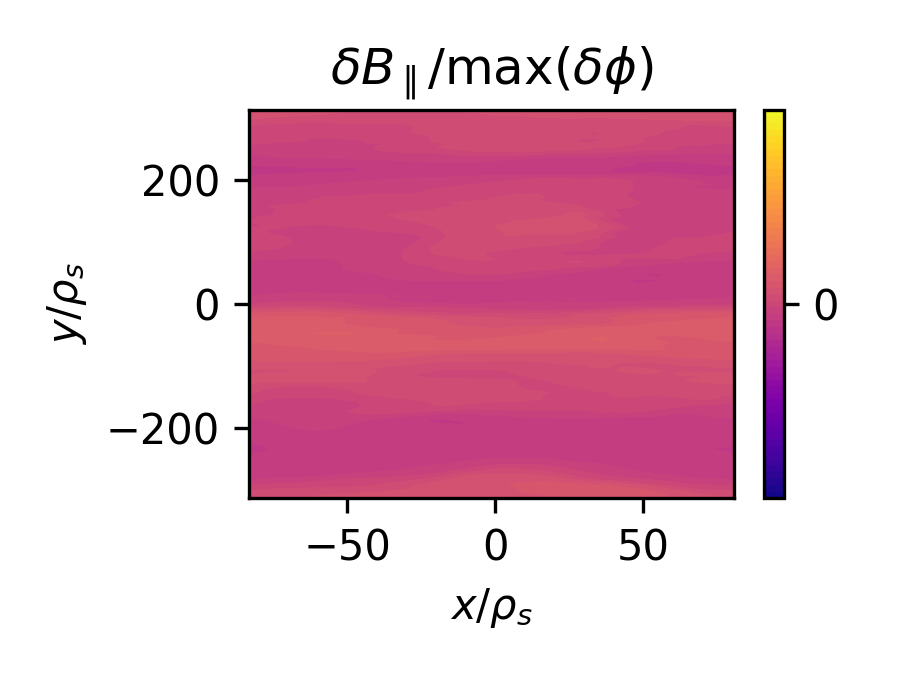}}\\

         \subfloat[$f_B = 0$ and MHD-1]{\includegraphics[width=0.3\textwidth]{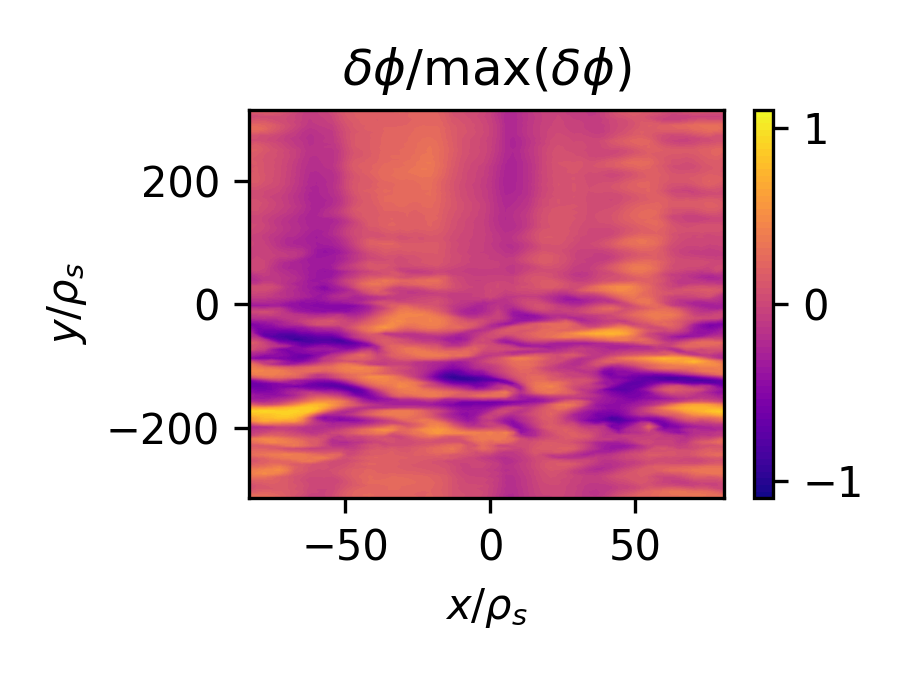}}\quad
   \subfloat[$f_B = 0$ and MHD-1]{\includegraphics[width=0.3\textwidth]{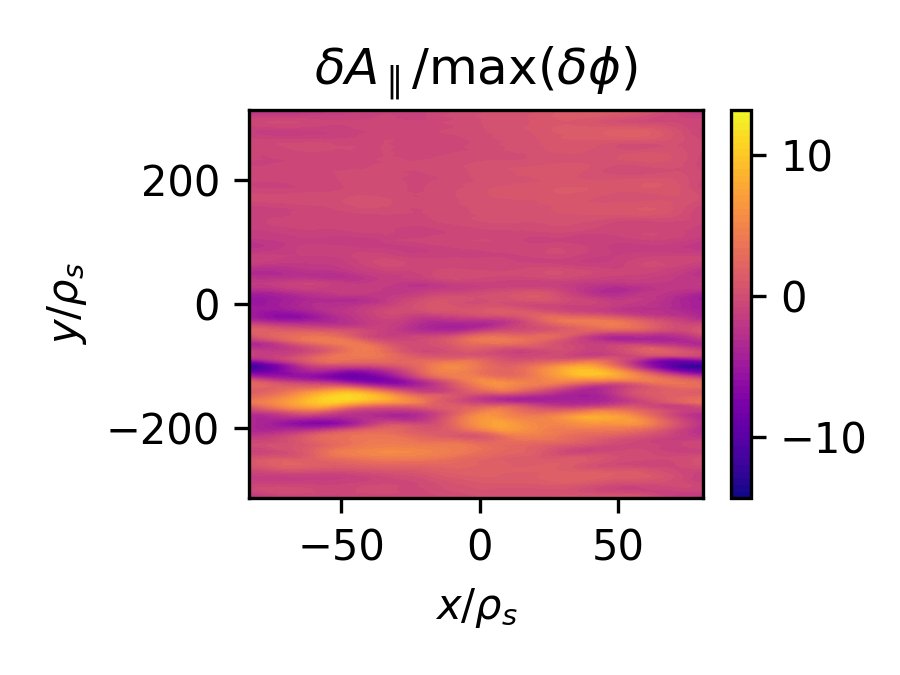}} \quad \subfloat[$f_B = 0$ and MHD-1]{\includegraphics[width=0.3\textwidth]{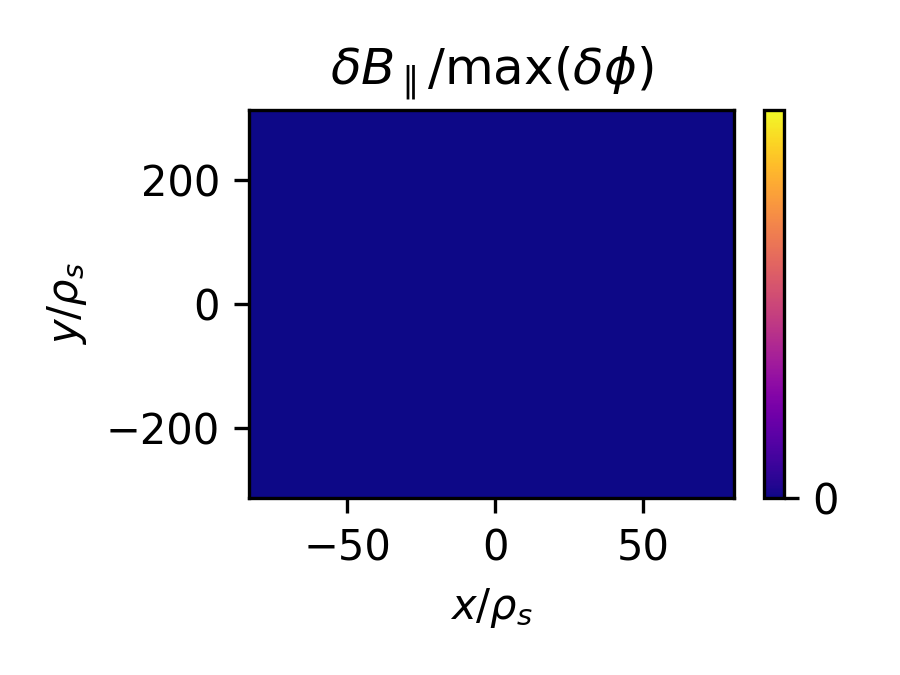}}\\

         \subfloat[$f_B = 0$ and $\beta^\prime = 0.4 \beta^\prime_{\mathrm{STEP}}$ ]{\includegraphics[width=0.3\textwidth]{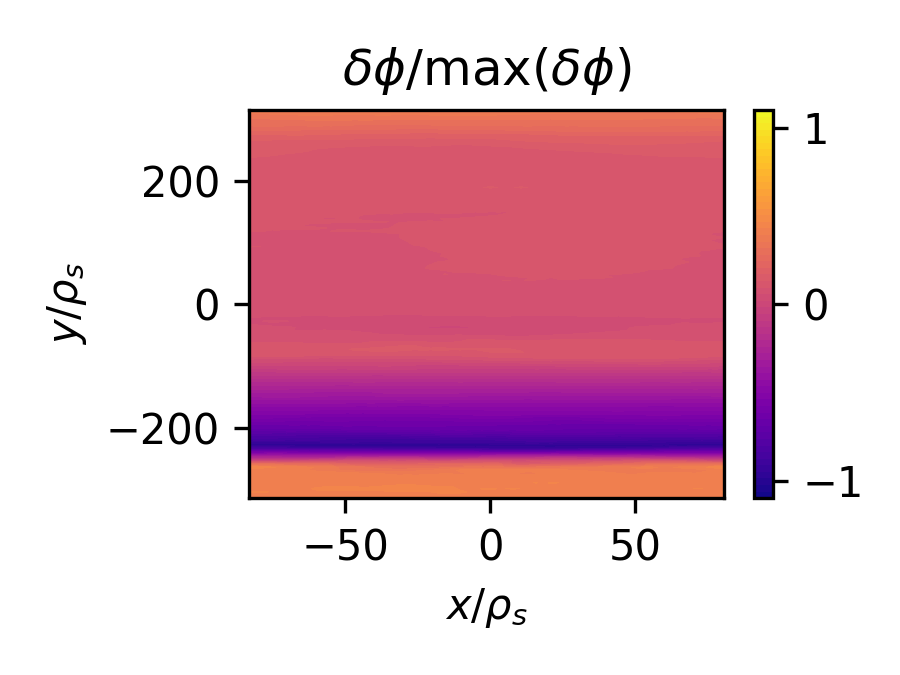}}\quad
   \subfloat[$f_B = 0$ and $\beta^\prime = 0.4 \beta^\prime_{\mathrm{STEP}}$ ]{\includegraphics[width=0.3\textwidth]{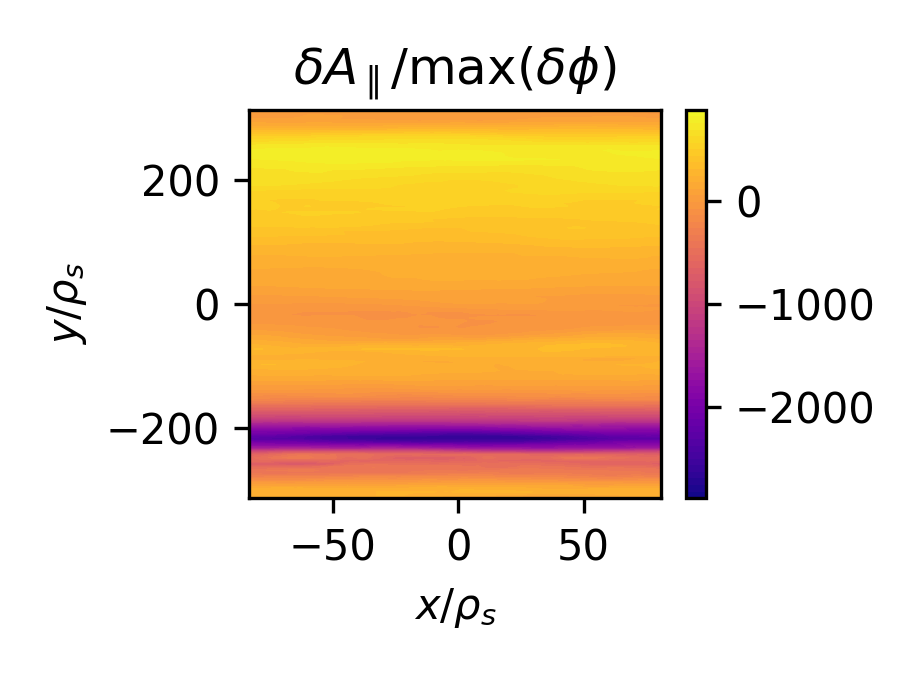}} \quad \subfloat[$f_B = 0$ and $\beta^\prime = 0.4 \beta^\prime_{\mathrm{STEP}}$ ]{\includegraphics[width=0.3\textwidth]{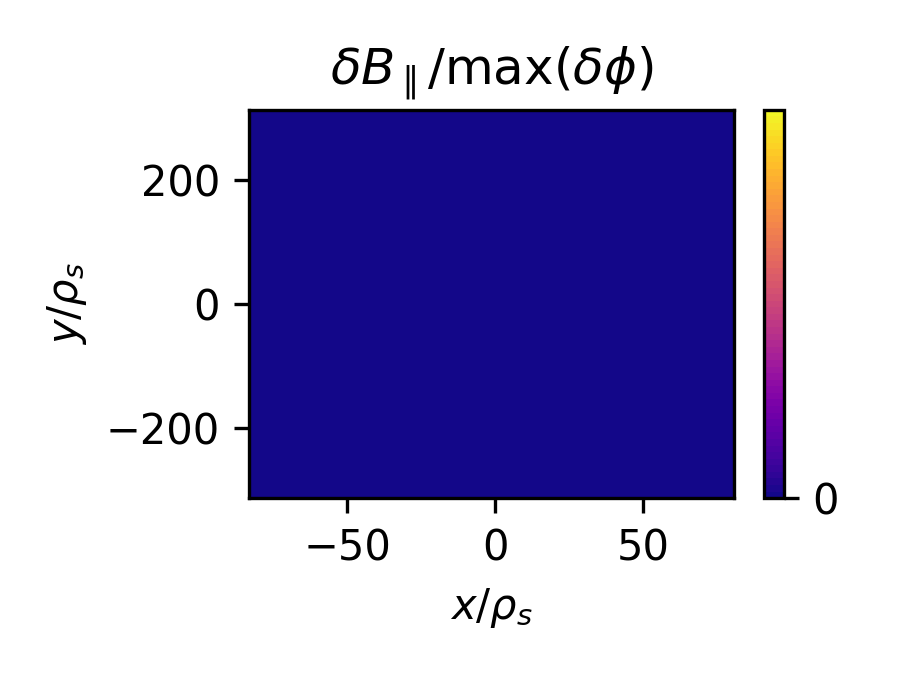}}\\


    \captionsetup{font=it}
     \caption{Snapshot contour plots at the outboard midplane ($\theta=0$) from the final time step of different nonlinear simulations discussed throughout the text: \emph{(\textbf{a, d, g})} $ e \delta \! \phi / (\rho_{\star} T_{e}),$ \emph{(\textbf{b, e, h})} $\delta \! A_\parallel / (\rho_\star \rho_s \B)$, and \emph{(\textbf{c, f, i})} $\delta \! B_\parallel / (\rho_\star T_{e} \B ).$ Each field is normalised to the maximum value of $ e \delta \! \phi / (\rho_{\star} T_{e})$ at the outboard midplane. {Simulations which return very large heat flux are typically associated with turbulent fields that exhibit radially-extended structures and reach large values of $(\delta \! A_{\parallel} T_{e}) / (\delta \! \phi e\rho_s \B).$} }
    \label{fig:turbulence_snapshots}
\end{figure}

\section*{References}
\bibliographystyle{unsrt}
\bibliography{bibliography}

\end{document}